\documentclass{llncs}
\usepackage{cite}
\usepackage{amsmath,amssymb,amsfonts}
\usepackage{graphicx}
\usepackage{caption}
\usepackage{subcaption}
\usepackage{rotating}
\usepackage{wrapfig}
\usepackage{smartdiagram}
\usepackage{float}
\usepackage{multirow}
\usepackage{bigfoot}
\DeclareNewFootnote[para]{default}
\begin{document}
\title{Multiperspective Conformance Analysis of Central Venous Catheter Installation Procedure}
\titlerunning{Multiperspective Conformance Analysis of CVC procedure}
\author{R.P. Jagadeesh Chandra Bose}
\institute{Accenture Labs, Bangalore, India\\
\email{jcbose@gmail.com}}
\maketitle

\begin{abstract}
Training and practice play a key role in a medical students' attainment of surgical procedural skills. It is beyond doubt that good skills correlate with better clinical outcomes and improved healthcare. Timely, holistic, and effective feedback provide a significant impetus to students acquiring skills with precision. In this paper, we analyze the activities performed by students while learning the central venous catheter installation procedure. We perform a holistic analysis, using {\it trace alignment, declarative conformance checking, data visualization, and statistical analysis techniques}, at different levels of abstraction on control-flow and time perspectives and provide insights at individual student level as well as across students. These insights can help students discover what they are doing right and where they are not and take corrective steps. Instructors can uncover common patterns and mistakes that students demonstrate and think of interventions in their teaching methodology.
\end{abstract}
\section{Introduction}
The first instance of the conformance checking challenge poses a problem of compliance analysis of students executing a medical procedure with respect to a well conceived protocol. The event log provided for the challenge captures the activities performed by students during their central venous catheter (CVC) installation procedure using ultrasound.  The event log contains execution traces of $10$ students during two rounds (PRE and POST), where PRE round corresponds to the first preliminary test on the procedure and POST round signifies the final assessment. The challenge specifies to analyze the data from two perspectives: (i) student and (ii) instructor.

Students might be interested in gaining insights on how they performed the procedure in the two rounds and what learning can they take. More specifically, the students might be interested in discovering insights such as {\it how compliant are they with respect to the protocol?}, {\it what sort of mistakes or deviations from the protocol do they do?}, {\it where do deviations/mistakes occur and how often do they occur?}, {\it whether their performance has improved in the POST round when compared with PRE?} etc. Instructors, on the other hand, are mostly interested in aggregated views of the performance of students, such as common mistakes manifested by students, similarities and differences exhibit by groups of students etc. Such learning can help them adapt their training/teaching methodology to enable students acquire their skills better.

In this paper, we attempt at analyzing this event log holistically from multiple perspectives. We focus on the control-flow and time aspects of the process execution. Control-flow conformance is analyzed using {\it trace alignment} \cite{tracealignment} and {\it Declarative conformance checking} \cite{burattin2016conformance}, both for individual students as well as across students. We also consider the analysis at the entire procedure level as well as at individual stages. From the time perspective, we analyzed the performance (processing time, turn-around time etc.). All of these analysis has been done for the PRE and POST rounds separately and the outcomes of the two are compared through {\it statistical techniques}. Our analysis unravels several common mistakes that students do while executing different stages of the procedure. We observed that the conformance and performance of students improved in their POST rounds when compared to their PRE rounds. Detailed observations are provided later in the paper.

The rest of the paper is organized as follows. Section~\ref{sec:datapreparation} presents some preprocessing and data preparation steps for our analysis. Section~\ref{sec:approach} elicits our approach of analysis and Section~\ref{sec:analysis} presents and discusses the results of our analysis. Finally, Section~\ref{sec:conclusions} concludes the paper.
\section{Data Preparation} \label{sec:datapreparation}
The event log provided for the challenge contains $20$ cases and $1394$ events over $29$ activities ($58$ event classes considering the start and complete event types). In this section, we discuss some of the preprocessing and data preparation aspects that we considered for our analysis. {\it We assume that the event log is noise free for our analysis.}
\subsection{Trace Name} For ease of referencing, we consider an alternate trace name (concept:name) for the various cases than what was provided in the log. The event log captured (plausibly) the video filename as the identifier of a case. We alter that to capture the resource and the round of examination to signify the trace name, i.e., we define the trace name as $<$RESOURCE$>$\_$<$ROUND$>$. \tablename~\ref{tab:tracenamemapping} depicts the mapping of trace names between the original log and our convention.
\begin{table}[!htb]
\centering
\caption{Mapping between the original concept:name attribute value and modified one based on resource and round attributes.}
\label{tab:tracenamemapping}
\begin{tabular}{l|l||l|l}
\hline
 \multicolumn{1}{c|}{Original } &\multicolumn{1}{c||}{ Modified } & \multicolumn{1}{c|} {Original } & \multicolumn{1}{c}{Modified }\\
 \multicolumn{1}{c|}{Trace Name}&\multicolumn{1}{c||}{ Trace Name} &\multicolumn{1}{c|} { Trace Name} & \multicolumn{1}{c}{ Trace Name}\\
\hline
\texttt{\detokenize{1539302414925-video_1.3_CVC}} & \texttt{\detokenize{R_13_1C_Pre}} & \texttt{\detokenize{1547683734202-video_1.c_CVC}} & \texttt{\detokenize{R_13_1C_Post}}\\
\texttt{\detokenize{1539303857517-video_1.4_CVC}} & \texttt{\detokenize{R_14_1D_Pre}} & \texttt{\detokenize{1547722738650-video_1.d_CVC}} & \texttt{\detokenize{R_14_1D_Post}}\\
\texttt{\detokenize{1539314415211-video_2.1_CVC}} & \texttt{\detokenize{R_21_1F_Pre}} &\texttt{\detokenize{1547693943698-video_1.f_CVC}} & \texttt{\detokenize{R_21_1F_Post}}\\
\texttt{\detokenize{1539316889981-video_3.1_CVC}} & \texttt{\detokenize{R_31_1G_Pre}} & \texttt{\detokenize{1547915248799-video_1.g_CVC}} & \texttt{\detokenize{R_31_1G_Post}}\\
\texttt{\detokenize{1539734942389-video_3.2_CVC}} & \texttt{\detokenize{R_32_1H_Pre}} & \texttt{\detokenize{1547698148253-video_1.h_CVC}} & \texttt{\detokenize{R_32_1H_Post}}\\
\texttt{\detokenize{1539737717686-video_3.3_CVC}} & \texttt{\detokenize{R_33_1L_Pre}} & \texttt{\detokenize{1547986862398-video_1.l_CVC}} & \texttt{\detokenize{R_33_1L_Post}}\\
\texttt{\detokenize{1539739275781-video_4.5_CVC}} & \texttt{\detokenize{R_45_2A_Pre}} & \texttt{\detokenize{1547997805806-video_2.a_CVC}} & \texttt{\detokenize{R_45_2A_Post}}\\
\texttt{\detokenize{1539740347920-video_4.6_CVC}} & \texttt{\detokenize{R_46_2B_Pre}} & \texttt{\detokenize{1548033283992-video_2.b_CVC}} & \texttt{\detokenize{R_46_2B_Post}}\\
\texttt{\detokenize{1539831132678-video_4.7_CVC}} & \texttt{\detokenize{R_47_2C_Pre}} & \texttt{\detokenize{1548034300723-video_2.c_CVC}} & \texttt{\detokenize{R_47_2C_Post}}\\
\texttt{\detokenize{1539832275246-video_4.8_CVC}} & \texttt{\detokenize{R_48_2D_Pre}} & \texttt{\detokenize{1548037113729-video_2.d_CVC}} & \texttt{\detokenize{R_48_2D_Post}}\\
\hline
\end{tabular}
\end{table}
\subsection{Event Timestamps}
The lifecycle and timestamp of an event are captured using the lifecycle:transition and  time:timestamp attributes respectively. The event log has two attributes \texttt{VIDEOSTART} and \texttt{VIDEOEND}, signifying the time when the activity started and ended in the video recording of students performing the CVC procedure. We consider these two attributes to define the \texttt{start} and \texttt{complete} timestamps of an activity rather than what was specified in the event log. The reasons for this are two fold:
\begin{itemize}
\item{\bf coarser granularity:} the granularity of the timestamps provided in the event log is coarse (the event log captures only at the minute granularity and ignores the seconds).
\item{\bf data quality:} certain data quality issues (incorrect values) are manifested in the event log. For example, for the trace \texttt{\detokenize{R_21_1F_Pre}}, the timestamp for event $44$, \texttt{Drop probe-complete}, is 11.10.2018 20:46:00.000 and the timestamp for event $47$, \texttt{Puncture-start}, is 11.10.2018 21.47:00.000. It is highly unlikely that there was a waiting time of $1$ hour between the two events. As another example, for the trace \texttt{\detokenize{R_13_1C_Post}}, the timestamp for event $22$, \texttt{Position probe-complete}, is 15.01.2019 19:39:00.000 while that of the next, event $23$, \texttt{Ultrasound configuration-start} is 16.01.2019 19:39:00.000. Here again it is highly unlikely that the gap between the two events is $1$ day.

    We believe that these are {\it unintentional} human errors crept while generating an event log by manually observing video recordings.
\end{itemize}
\subsection{Protocol Trace}
\begin{figure}[!htb]
\centering
\includegraphics[width=0.99\textwidth]{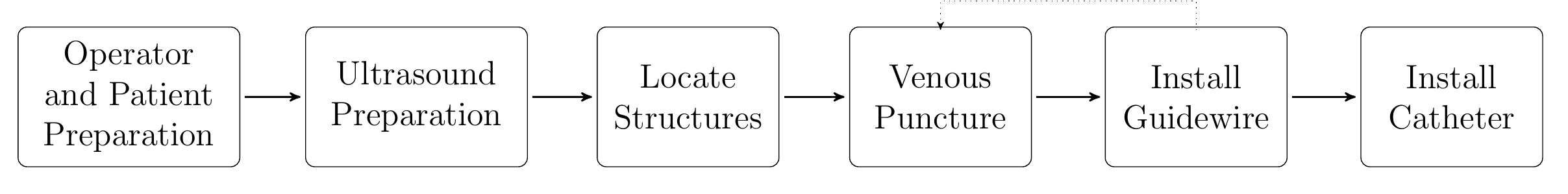}
\caption{Reference CVC installation procedure at stage level.}
\label{fig:protocolstagelevel}
\end{figure}

\begin{figure}[!htb]
\centering
\includegraphics[width=0.99\textwidth]{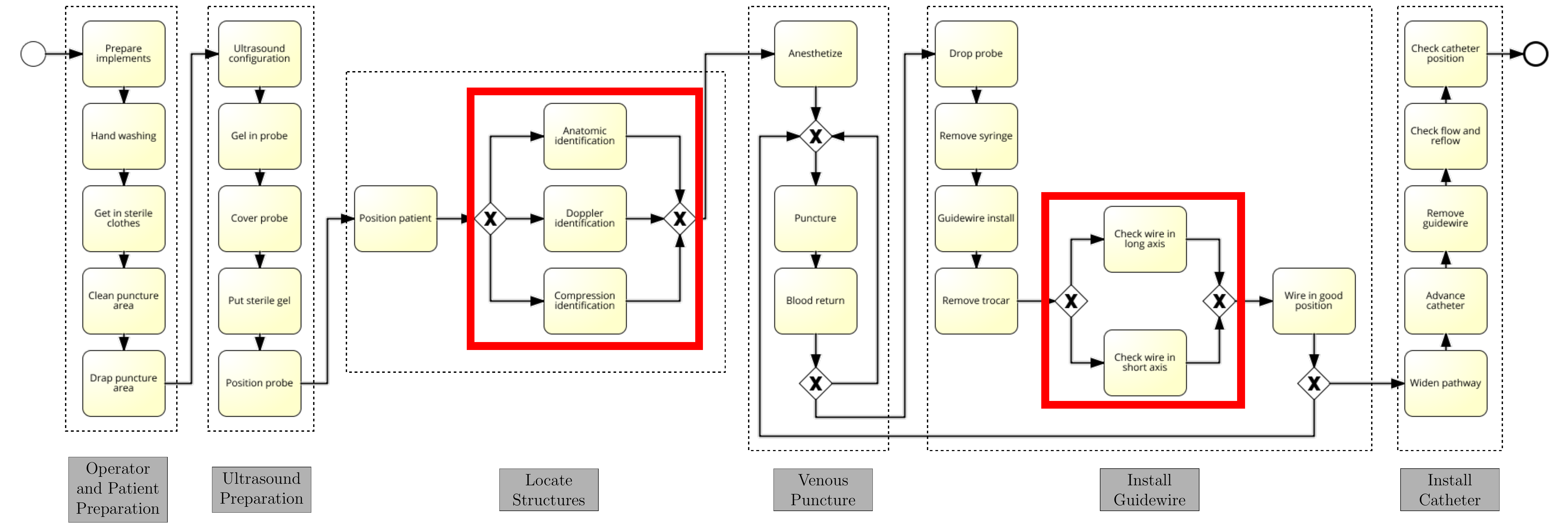}
\caption{Reference CVC installation procedure at activity level. The model is annotated with stages.}
\label{fig:bpmnmodel}
\end{figure}

Along with the dataset, the challenge provided a reference process model capturing the consensus procedure to be followed by medical practitioners during central venous catheter installation with ultrasound. The protocol at a very high-level signifies a sequential process as illustrated in \figurename~\ref{fig:protocolstagelevel} comprising of the following stages: \texttt{operator and patient preparation}, \texttt{ultrasound preparation}, \texttt{locate structures}, \texttt{venous puncture}, \texttt{install guidewire}, and \texttt{install catheter}. Each stage is further defined over a set of fine-grained activities. With the exception of few activities, the overall process is sequential. \figurename~\ref{fig:bpmnmodel} depicts the BPMN model of the CVC procedure annotated with the different stages. There are two {\it exclusive choice} constructs in the model (highlighted with solid rectangles in \figurename~\ref{fig:bpmnmodel}): (i) in the {\sf locate structures} stage where one of {\it anatomic identification}, {\it compression identification}, and {\it doppler identification} is to be executed and (ii) in the {\sf install guidewire} stage where either of {\it check wire in long axis} or {\it check wire in short axis} is to be executed.
We used the reference model and defined a {\bf protocol} trace. Since we have only the control-flow perspective of the reference model, we generate the protocol trace with only one event type, viz., complete event type. The protocol trace will be added to the event log for conformance analysis (using trace alignment \cite{tracealignment}). The trace name (concept:name) of the protocol trace is set to \texttt{Protocol Trace}. For the two exclusive choice constructs, since only one among the activities involved in the choice construct has to be executed, we defined the two following activities in the protocol trace: (i) \textsf{A\textunderscore D\textunderscore C identification} and (ii) \textsf{Check wire in l\textunderscore s axis}. Furthermore, although there are two loop constructs in the reference model, we consider only one iteration of the loop in the protocol trace. 
\subsection{Stage As Activity Transformation}
In order to do performance analysis at stage level (e.g., time spent in various stages), we applied the following transformation where each fine-grained activity is replaced with its corresponding stage. The event log obtained upon this transformation contains $20$ cases with $1394$ events over $12$ event classes (corresponding to the {\it start} and {\it complete} event types of the six stages).
\subsection{Stage Abstraction}
In order to analyze how students perform during various stages of the procedure, we create sub-logs for each stage. Each trace of the original event log is sliced according to their stage, which forms a case of the sub-log. For example, all events pertaining to the operator and patient preparation in a case are created as a case for that stage's sub-log. \tablename~\ref{tab:sublogcharacteristics} depicts the characteristics (the number of events and event classes) of the sub-logs for the six stages.
\begin{table}
\centering
\caption{Event log characteristics of sub-logs for the various stages.}
\label{tab:sublogcharacteristics}
\begin{tabular}{l|c|c}
\hline
\hline
Stage & \#Event Classes & \#Events \\
\hline
\hline
Operator and Patient Preparation & $10$ & $322$ \\
Ultrasound Preparation & $10$ & $238$\\
Locate Structures & $8$ & $114$ \\
Venous Puncture & $6$ & $156$ \\
Install Guidewire & $14$ & $348$ \\
Install Catheter & $10$ & $216$ \\
\hline
\hline
\end{tabular}
\end{table}
\subsection{Complete Only Events}
For conformance analysis with respect to control-flow of various activities as specified in the reference model, we consider only the events from the event log whose event type is complete.
\section{Approach} \label{sec:approach}
\begin{figure}[!htb]
\centering
\includegraphics[width=0.8\textwidth]{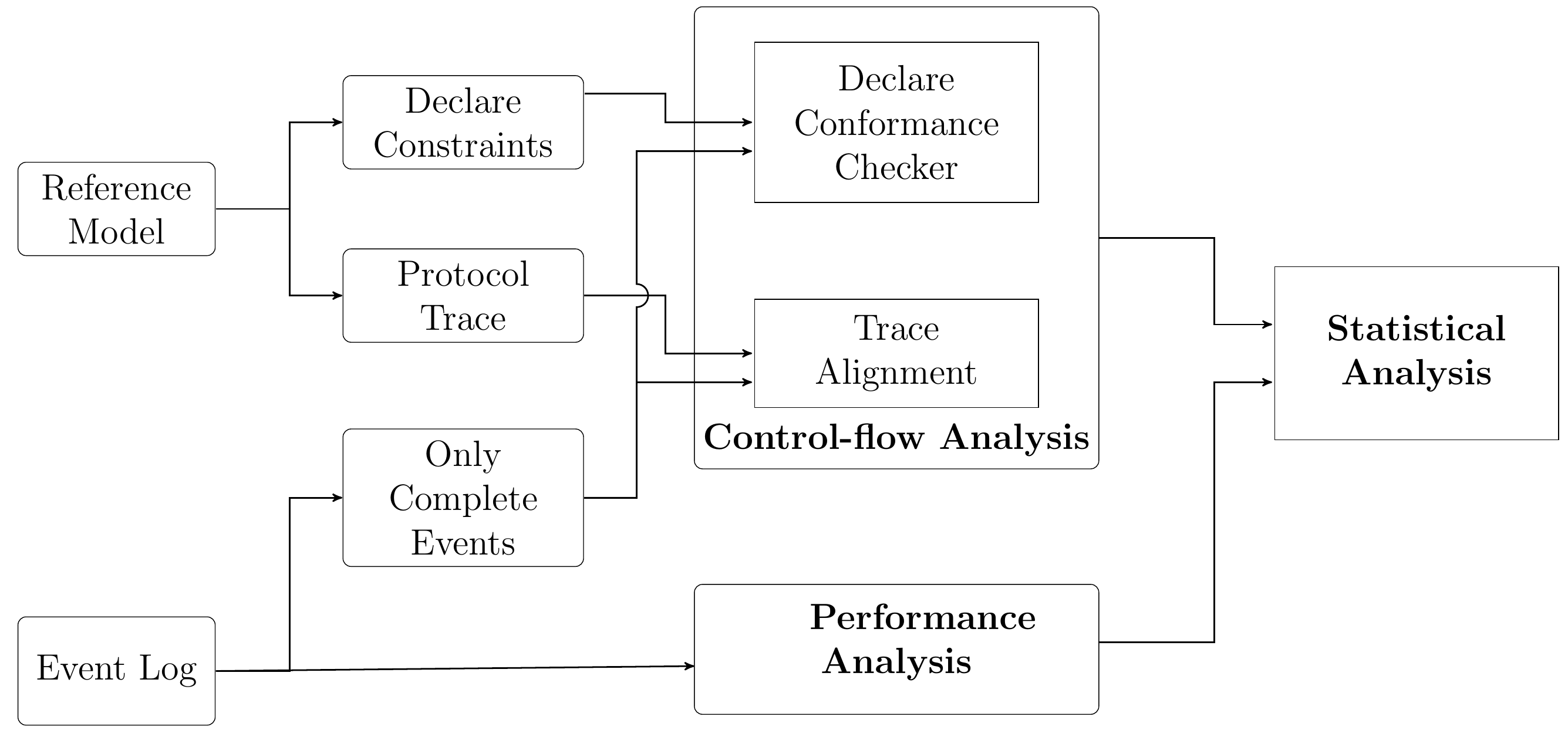}
\caption{Approach adopted in this paper for multiperspective conformance analysis.}
\label{fig:approach}
\end{figure}

We adopt the approach illustrated
 in \figurename~\ref{fig:approach} in this paper. We analyze the event log on two different aspects, viz., control-flow and time. For the control-flow analysis, we check for the compliance of the event log w.r.t the reference model. We consider two approaches for control-flow conformance analysis, (i) based on declarative models and (ii) based on trace alignment (we discuss on these approaches later in this section). First, we define a set of declarative constraints based on the reference model and consider only the complete events of the event log for declarative conformance checking. For the trace alignment, we derive a {\it protocol trace} as discussed in the previous section and align it with the complete only event log. We perform performance analysis on time such as processing time, turnaround time, and idle time analysis using the event log. We do these analysis at two different levels of granularity, one for the entire procedure and another for the different stages of the procedure considering individual students separately as well as across all students. Furthermore, we analyze the event log for the PRE and POST rounds separately and do a statistical analysis to assess whether any significant differences exist between the two rounds on the compliance and performance aspects.
\subsection{Declare Modeling and Analysis}
Healthcare processes are typically considered to be flexible. {\it Imperative modeling} paradigms, like the reference model specified in BPMN, wherein the process models capture all allowed activity flows are found to be shortcoming. Several studies recommend the use of {\it declarative modeling} to capture flexible processes \cite{pesic2007constraint,reichert2012enabling}. A declarative model captures a process under an ``open world" assumption, where everything is allowed unless it is explicitly forbidden by a rule/constraint. For example, DECLARE \cite{pesic2007declare} is a declarative process modeling language where a process is specified
via a set of constraints between activities, which must be satisfied by every execution of the process. \tablename~\ref{tab:declaretemplates} specifies some of the Declare templates and their interpretation. We use these templates to model some of the critical aspects\footnote{We consider these to be critical elements of the process based on intuition. The idea here is to show the approach; domain experts can specify the constraints of importance and the approach can be adopted on them.} of the CVC procedure and analyze how conformant are the student executions w.r.t this model.

\begin{table}[!htb]
\centering
\caption{Declare templates and their interpretation.}
\label{tab:declaretemplates}
\begin{tabular}{l|p{0.7\textwidth}}
\hline
Relation & Interpretation \\
\hline
\hline
Precedence(A, B) & ``B" has to be preceded by ``A". ``B" can happen only after ``A" had happened. \\
\hline
Response(A,B) & Whenever activity ``A" is executed, activity ``B" has to be eventually executed afterwards. \\
\hline
Alternate Response (A,B) & After each ``A" is executed at least one ``B" is executed. Another ``A" can be executed again only after the first ``B". \\
\hline
Exclusive Choice 1~of~3(A,B,C) & Only one activity (from A, B and C) has to be executed. \\
\hline
Exactly1(A) & A has to happen exactly once.\\
\hline
\end{tabular}
\end{table}

\tablename~\ref{tab:declaremodel} specifies $10$ DECLARE constraints of the CVC procedure. These constraints were chosen based on our intuition. For example, it is logical to expect that one washes their hands before touching/putting on sterile clothes. As another example, it is natural to expect that a patient is anesthetized only once (as an over dose of anesthetization can be dangerous). This is captured using the Exactly 1 constraint on this activity.
\begin{table}[!htb]
\centering
\caption{Declare constraints for the CVC procedure.}
\label{tab:declaremodel}
\begin{tabular}{l|p{0.9\textwidth}}
\hline
\hline
S.No & DECLARE Constraint \\
\hline
\hline
1 & Precedence(Hand washing, Get sterile clothes) \\
\hline
\multirow{3}{*}{2} & Precedence(Position patient, Anatomic identification) \\
 & Precedence(Position patient, Doppler identification) \\
 & Precedence(Position patient, Compression identification)\\
\hline
3 & Exclusive Choice 1~of~3 (Anatomic identification, Doppler identification, Compression identification) \\
\hline
4 & Exactly 1 (Anesthetize) \\
\hline
5 & Precedence(Anesthetize,Puncture)\\
\hline
6 & Alternate Response(Puncture,  Blood return)\\
\hline
7 & Precedence(Remove Trocar, Wire in good position) \\
\hline
8 & Precedence(Wire in good position, Advance catheter)\\
\hline
9 & Response(Remove guidewire, Check flow and reflow)\\
\hline
10 & Response(Install guidewire,  Remove guidewire) \\
\hline
\end{tabular}
\end{table}
\subsection{Trace Alignment}
Trace alignment has been proposed as a powerful technique for process diagnostics \cite{tracealignment}. The goal of trace alignment is to align the traces in such a way that event logs can be explored easily.  By aligning traces we can see the common and frequent behavior, and distinguish this from the exceptional behavior. Trace alignment assists in answering questions such as {\it what is the most common (likely) process behavior that is executed?}, {\it where do process instances deviate and what do they have in common?},  {\it are there any common patterns of execution in the traces?} etc. We align the traces capturing the students execution of the CVC procedure with the protocol trace to find commonalities and deviations (non-conformance). Trace alignment works by transforming traces in an event log into encoded character sequences.
\tablename~\ref{tab:activitycharencoding} depicts the character encoding of various activities in the event log. For example, all occurrences of the activity \texttt{\detokenize{Advance catheter-complete}} in the event log are represented by the alphabet \texttt{\detokenize{a}}.
\begin{table}[!htb]
\centering
\caption{Activities and their character encodings used in trace alignment. The event type of all the activities is complete.}
\label{tab:activitycharencoding}
\begin{tabular}{c|l||c|l}
\hline
\hline
 \multicolumn{1}{c|}{Char } &\multicolumn{1}{c||}{ Activity } & \multicolumn{1}{c|} {Char } & \multicolumn{1}{c}{Activity }\\
 \multicolumn{1}{c|}{Encoding}&\multicolumn{1}{c||}{ Name} &\multicolumn{1}{c|} { Encoding} & \multicolumn{1}{c}{ Name}\\
\hline
\hline
\texttt{\detokenize{a}} & \texttt{\detokenize{Advance catheter}} & \texttt{\detokenize{q}} & \texttt{\detokenize{Get in sterile clothes}}\\
\texttt{\detokenize{b}} & \texttt{\detokenize{Anatomic identification}} & \texttt{\detokenize{r}} & \texttt{\detokenize{Guidewire install}}\\
\texttt{\detokenize{c}} & \texttt{\detokenize{Anesthetize}}& \texttt{\detokenize{s}} & \texttt{\detokenize{Hand washing}}\\
\texttt{\detokenize{d}} & \texttt{\detokenize{Blood return}}& \texttt{\detokenize{t}} & \texttt{\detokenize{Position patient}}\\
\texttt{\detokenize{e}} & \texttt{\detokenize{Check catheter position}}& \texttt{\detokenize{u}} & \texttt{\detokenize{Position probe}}\\
\texttt{\detokenize{f}} & \texttt{\detokenize{Check flow and reflow}}& \texttt{\detokenize{v}} & \texttt{\detokenize{Prepare implements}}\\
\texttt{\detokenize{g}} & \texttt{\detokenize{Check wire in l_s axis}}& \texttt{\detokenize{w}} & \texttt{\detokenize{Puncture}}\\
\texttt{\detokenize{h}} & \texttt{\detokenize{Check wire in long axis}}& \texttt{\detokenize{x}} & \texttt{\detokenize{Put sterile gel}}\\
\texttt{\detokenize{i}} & \texttt{\detokenize{Check wire in short axis}}& \texttt{\detokenize{y}} & \texttt{\detokenize{Remove guidewire}}\\
\texttt{\detokenize{j}} & \texttt{\detokenize{Clean puncture area}}& \texttt{\detokenize{z}} & \texttt{\detokenize{Remove syringe}}\\
\texttt{\detokenize{k}} & \texttt{\detokenize{Compression identification}}& \texttt{\detokenize{A}} & \texttt{\detokenize{Remove trocar}}\\
\texttt{\detokenize{l}} & \texttt{\detokenize{Cover probe}} & \texttt{\detokenize{B}} & \texttt{\detokenize{Ultrasound configuration}}\\
\texttt{\detokenize{m}} & \texttt{\detokenize{Doppler identification}} & \texttt{\detokenize{C}} & \texttt{\detokenize{Widen pathway}}\\
\texttt{\detokenize{n}} & \texttt{\detokenize{Drap puncture area}}& \texttt{\detokenize{D}} & \texttt{\detokenize{Wire in good position}}\\
\texttt{\detokenize{o}} & \texttt{\detokenize{Drop probe}}& \texttt{\detokenize{E}} & \texttt{\detokenize{A_D_C identification}}\\
\texttt{\detokenize{p}} & \texttt{\detokenize{Gel in probe}}& & \\
\hline
\hline
\end{tabular}
\end{table}

\section{Analysis Results and Discussion}\label{sec:analysis}
In this section, we present our results of analysis using the approach elicited in the previous section. We analyze the event log from two different perspectives: (i) individual student and (ii) instructor. We highlight some of the questions that each  might be interested in and address them through our analysis.
\subsection{Student Perspective} The focus of analysis here is on the individual students. The students are interested in gaining insights on how they performed the procedure in the two rounds and what learning can they take. More specifically, the students might be interested in knowing:
\begin{itemize}
\item how compliant are they with respect to the protocol?
\item what sort of mistakes or deviations from the protocol do they do?
\item where do deviations/mistakes occur?
\item how often do deviations/mistakes occur?
\item whether their performance has improved in the POST round when compared with PRE?
\item whether they struggle in performing some activities, if so, where?
\end{itemize}

The student might be interested in their performance both at the overall procedure level as well as at a stage level. We focus on two aspects in our analysis: (i) control-flow and (ii) time. We analyze the control-flow aspects of a student's execution of the procedure using trace alignment. We consider only the complete events of activities in the event log for trace alignment.

\subsubsection{Entire Procedure Level}
\figurename~\ref{fig:R131CTraceAlignmentPre} depicts the alignment of the trace pertaining to resource \texttt{\detokenize{R_13_1C}} in the PRE round w.r.t the protocol trace. Deviations (or non-conformant behavior) from the protocol trace are manifested as gaps `-' in the alignment. Common activities/patterns are captured as well conserved regions in the alignment. There are a total of $19$ deviations, which are listed as follows:
\begin{figure}[!htb]
\centering
\includegraphics[width=0.99\textwidth]{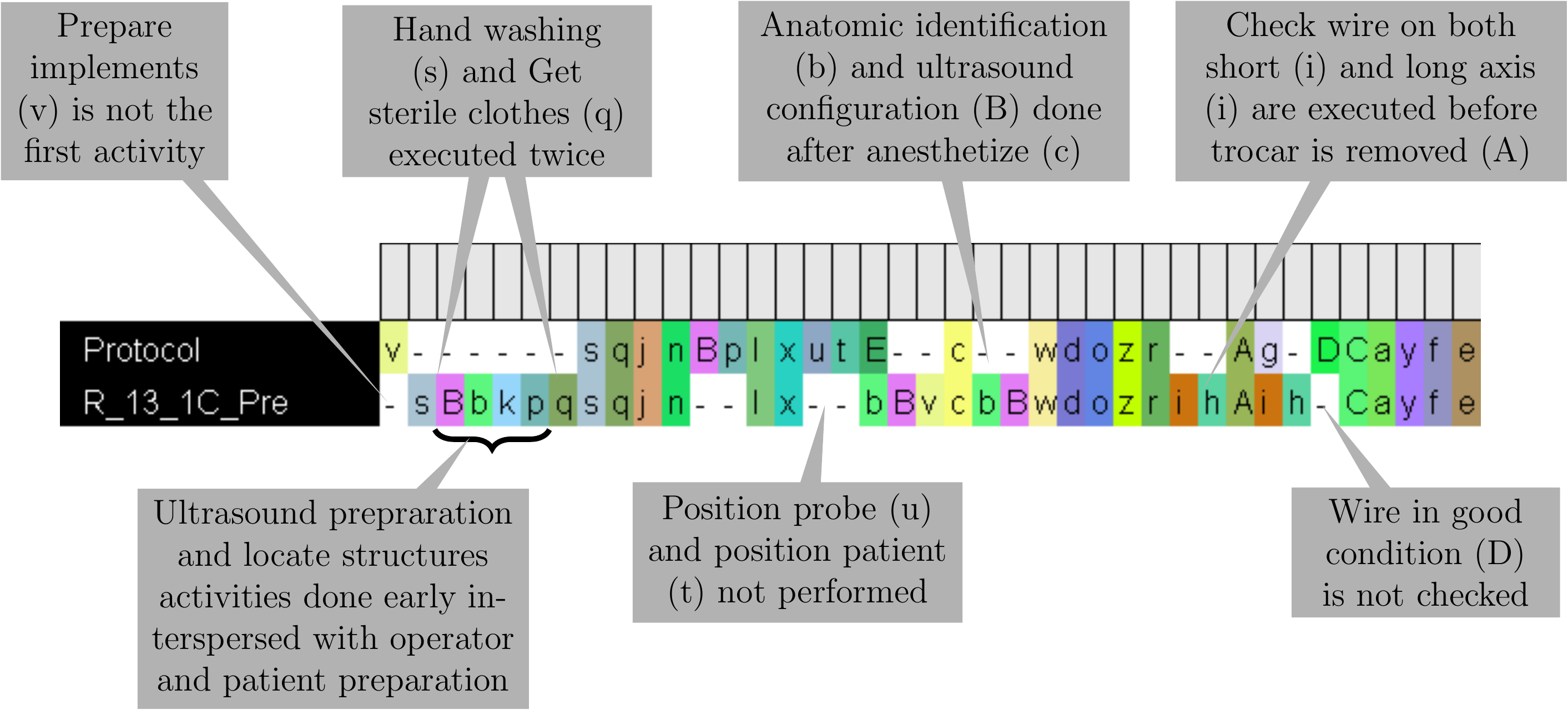}
\caption{Trace alignment of the PRE round trace with the protocol for the student \texttt{\detokenize{R_13_1C}}.}
\label{fig:R131CTraceAlignmentPre}
\end{figure}
\begin{itemize}
 \item the pre trace does not start with the {\sf Prepare Implements} activity (represented by `v' in the alignment). In fact, this activity was done quite late in the procedure just before {\sf Anesthetize} (`c')
 \item while the protocol specifies the execution of all activities pertaining to the operator and patient preparation before moving on to the other stages, \texttt{\detokenize{R_13_1C}} jumped to ultrasound preparation activities (`B' and `p') and locate structures (`b' and `k') before completing operator and patient preparation
 \item while the protocol specifies that hand washing (`s') and getting sterile clothes (`q') need to be executed once, \texttt{\detokenize{R_13_1C}} performed these activities twice
 \item \texttt{\detokenize{R_13_1C}} did not perform {\sf Position probe} activity (`u') under ultrasound preparation and {\sf Position patient} (`t') under locate structures
 \item while the protocol specifies that only one of anatomic, compression, or doppler identification has to be executed, we notice both anatomic, `b', and compression identification, `k', executed in this trace. Note that the activity `b' is aligned with activity `E' from the protocol trace. As mentioned earlier, the protocol trace has the representative \texttt{\detokenize{A_D_C identification}} to capture the choice and any of the three activity manifestations is aligned with this. In addition, the anatomic identification activity `b' is executed thrice in the trace where as the protocol suggests only one
 \item Ultrasound configuration (`B') and anatomic identification (`b') are performed even after Anesthetize (`c')
 \item \texttt{\detokenize{R_13_1C}} performed the check wire on both the long and short axis (`i' and `h') unlike the protocol which suggests either of them. Furthermore, the protocol mandates that either of these activities be executed after remove trocor (`A') but the student performed both these activities twice, once before the trocar is removed and once after. Note that the protocol had a representative of this choice construct as the activity \texttt{\detokenize{Check wire l_s axis}}, denoted by activity `g' in the alignment. `g' is aligned to either of `i' or `h' in the alignment to signify their equivalence
 \item \texttt{\detokenize{R_13_1C}}  did not perform the activity {\sf Wire in good condition} (reflected as a missing activity `D') before the install catheter stage
\end{itemize}

\begin{figure}[!htb]
\centering
\includegraphics[width=0.99\textwidth]{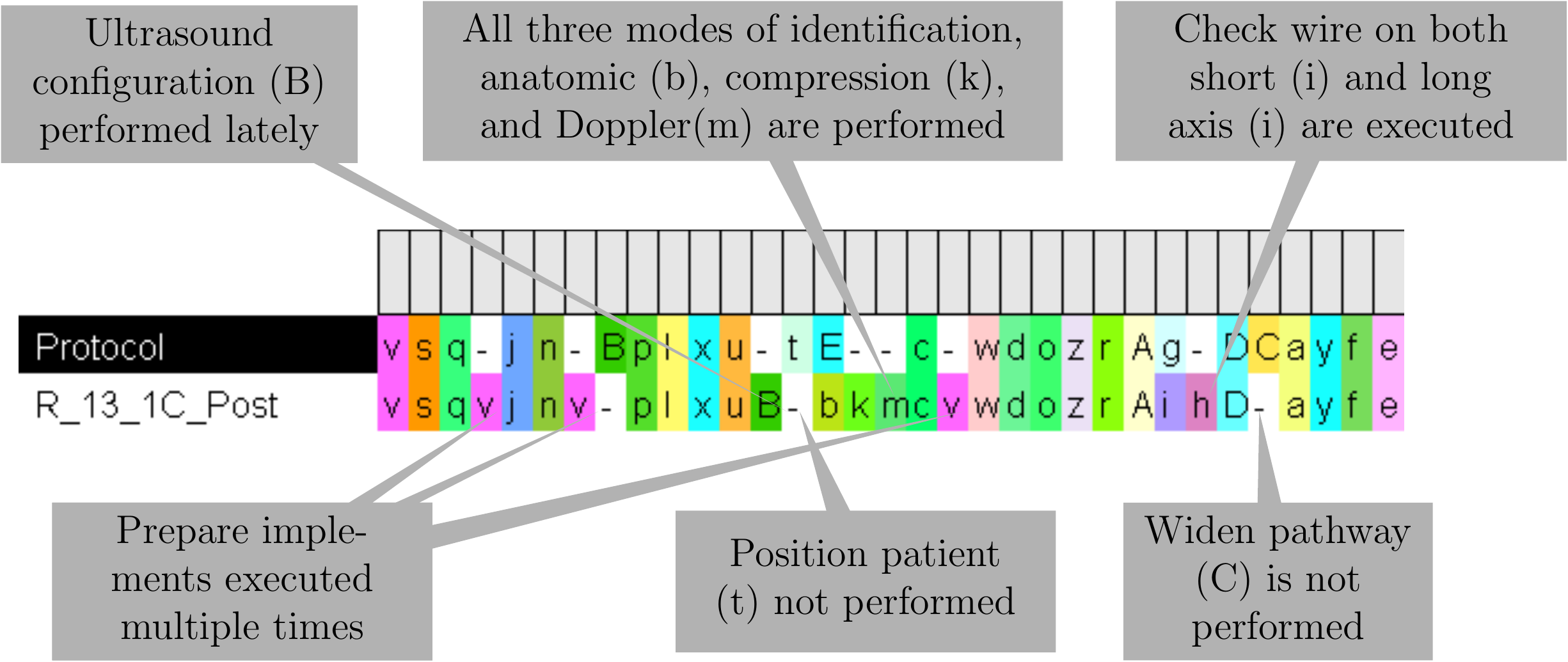}
\caption{Trace alignment of the POST round trace with the protocol for the student \texttt{\detokenize{R_13_1C}}.}
\label{fig:R131CTraceAlignmentPost}
\end{figure}
\begin{figure}[!htb]
\centering
\includegraphics[width=0.99\textwidth]{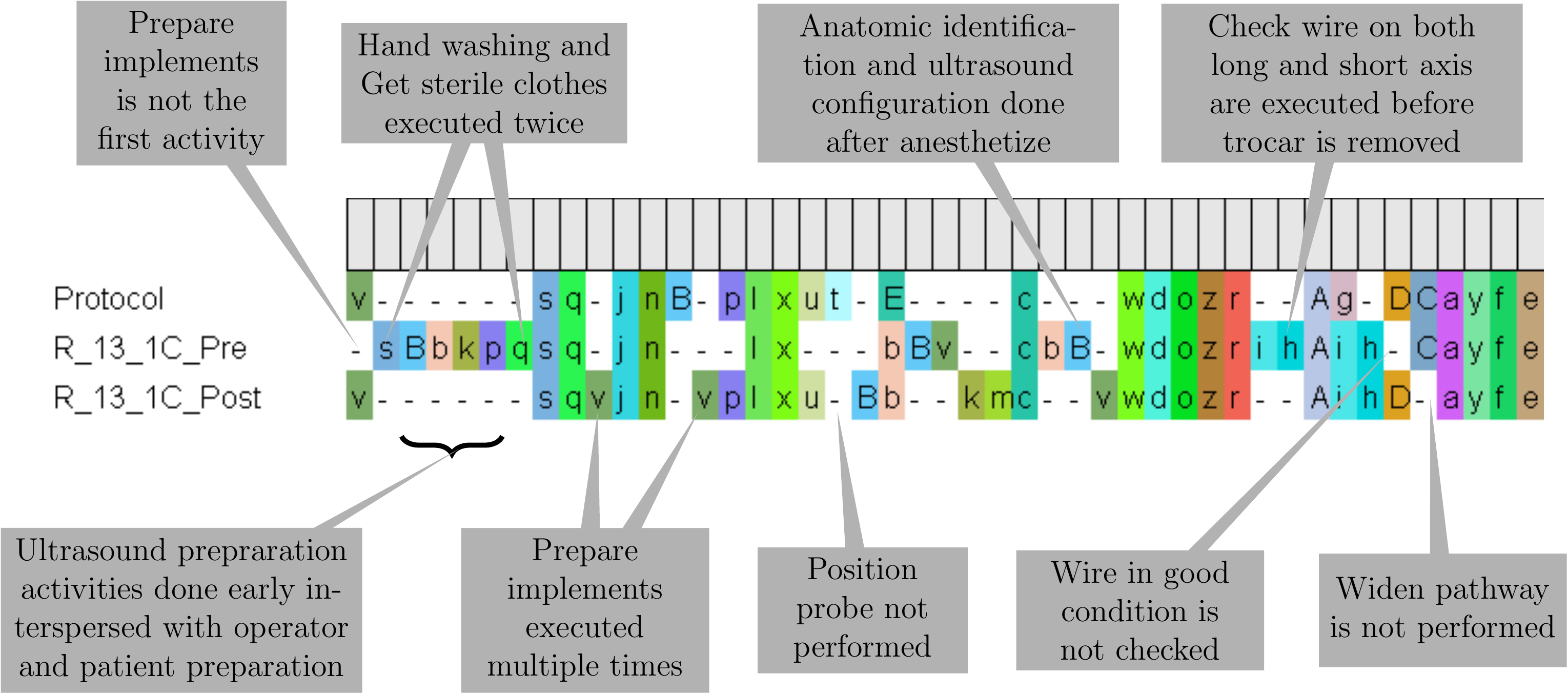}
\caption{Trace alignment of the PRE and POST round traces with the protocol for the student \texttt{\detokenize{R_13_1C}}.}
\label{fig:R131CTraceAlignmentBoth}
\end{figure}

\figurename~\ref{fig:R131CTraceAlignmentPost} depicts the alignment of the protocol trace with the procedure followed by \texttt{\detokenize{R_13_1C}} during the POST round. There are a total of $10$ deviations, as listed below.
\begin{itemize}
\item the activity Prepare implements (`v') has been executed four times, out of which three instances were executed within the operator and patient preparation stage, while one instance was executed after Anesthetize
\item the activity Ultrasound configuration (`B') has been executed as the last activity of the Ultrasound preparation stage whereas the protocol specifies it to be the first activity
\item the activity Position patient (`t') has not been executed
\item all three modes of identification, viz., anatomic (`b'), compression (`k'), and doppler (`m'), has been performed in the locate structures stage while the protocol specifies one of the three to be executed
\item in the install guidewire stage, check wires on both long (`h') and short (`i') axis has been performed where as the protocol specifies either of them to suffice
\item the activity Widen pathway (`C') has not been executed in the install catheter stage
\end{itemize}
\figurename~\ref{fig:R131CTraceAlignmentBoth} depicts the alignment of the protocol trace with the procedures followed by \texttt{\detokenize{R_13_1C}} during both the PRE and POST rounds. We could clearly see that the procedure followed in the POST round to be more aligned/compliant with the protocol trace (also evident with 10 vs. 19 deviations). While in the PRE round, activities belonging to different stages were interspersed (e.g., ultrasound preparation/locate structures activities executed before operator and patient preparation is completed), such a behavior is almost absent (with the exception of one instance of Prepare implements executed after Anesthetize). Similarly, repeated executions of activities has also reduced in the POST round. A common deviation that is persistent in both the PRE and POST rounds is w.r.t the activities involved in the two choice constructs. In both instances while the protocol specifies executing one of the activities involved suffices, more than one activity was performed by the student.

\subsubsection{Individual Stages Level}
\begin{figure}[!htb]
\centering
\begin{subfigure}[t]{0.45\textwidth}
\centering
\includegraphics[width=0.99\linewidth]{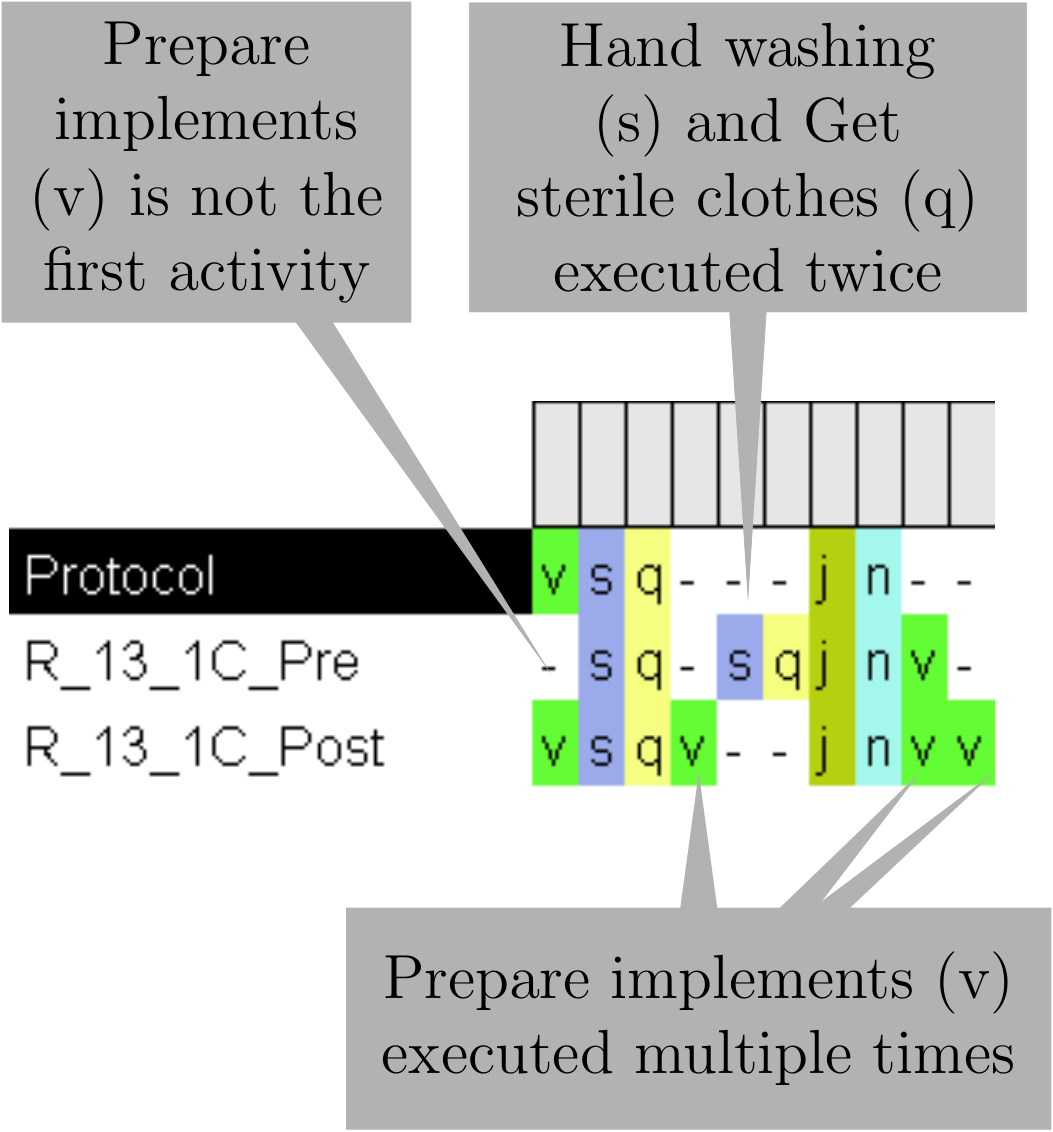}
\caption{Operator and patient preparation}
\label{fig:R_13_1C_OperatorAndPatientPreparation}
\end{subfigure}
\hfill
\begin{subfigure}[t]{0.4\textwidth}
\centering
\includegraphics[width=0.99\linewidth]{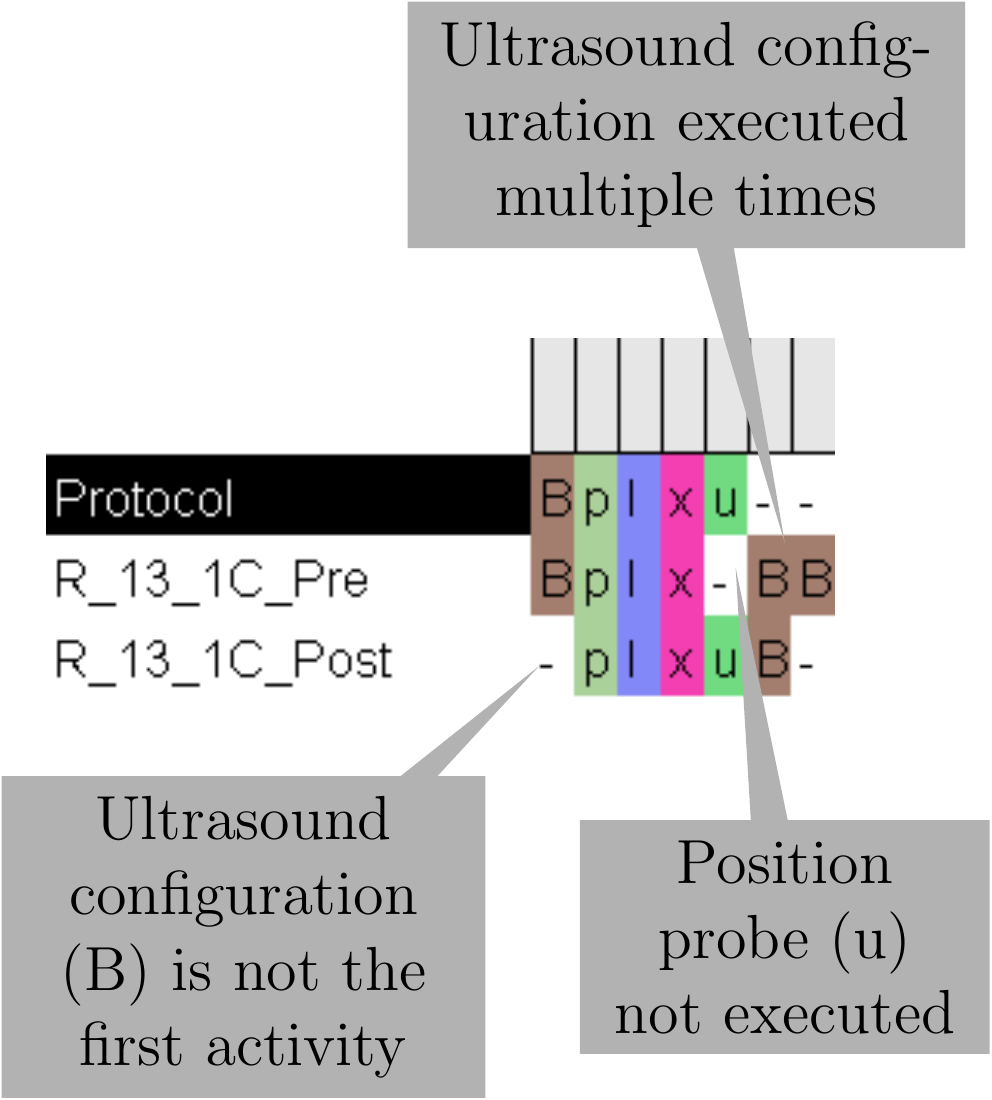}
\caption{Ultrasound preparation}
\label{fig:R_13_1C_UltrasoundPreparation}
\end{subfigure}
\caption{Trace alignment of the operator and patient preparation and ultrasound preparation sublogs for the student \texttt{\detokenize{R_13_1C}}.}
\label{fig:R_13_1C_TraceAlignmentSubLogOPUP}
\end{figure}

\begin{figure}[!htb]
\centering
\begin{subfigure}[t]{0.33\textwidth}
\includegraphics[width=0.99\linewidth]{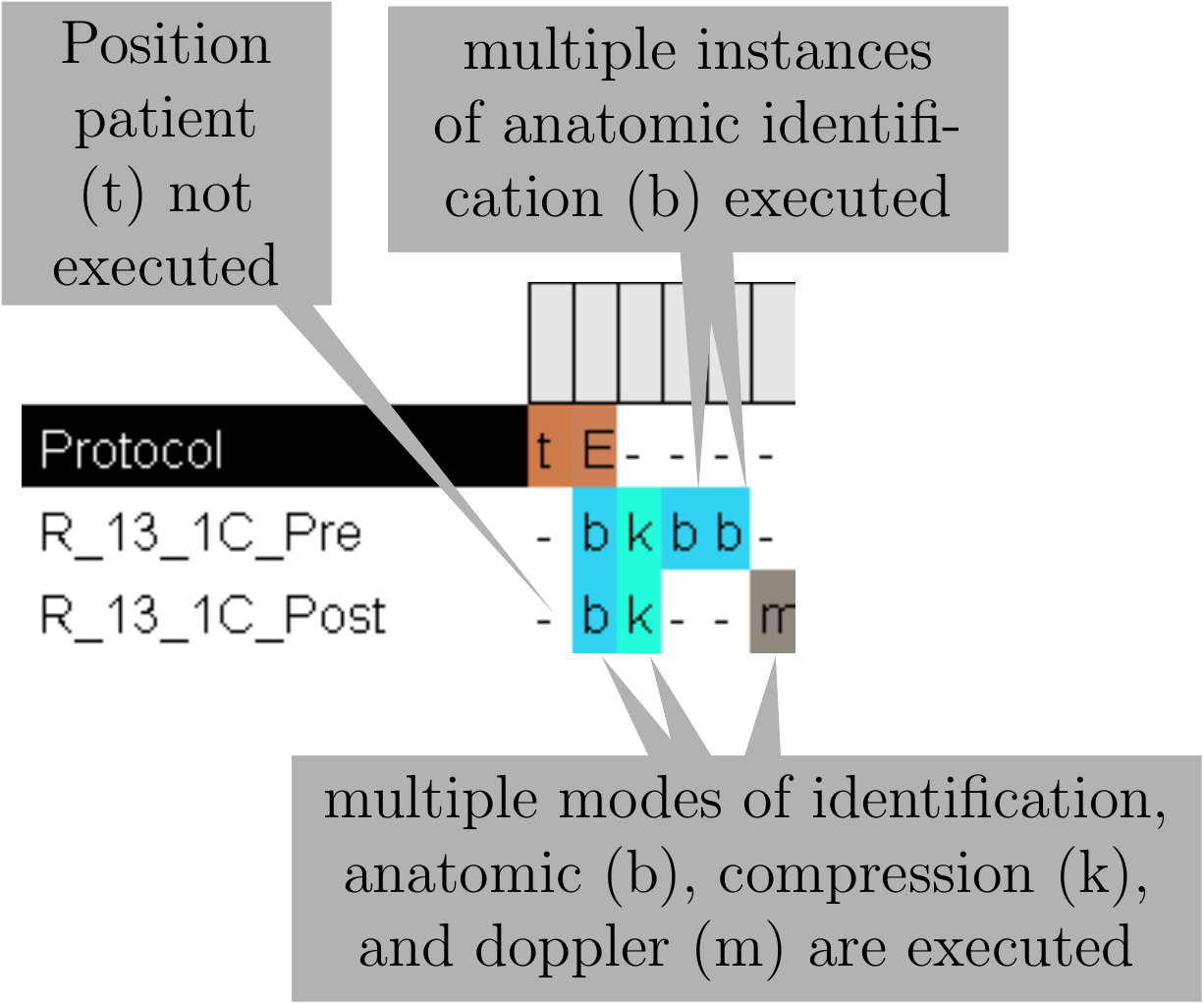}
\caption{Locate structures}
\label{fig:R_13_1C_locatestructures}
\end{subfigure}
\hfill
\begin{subfigure}[t]{0.33\textwidth}
\includegraphics[width=0.99\linewidth]{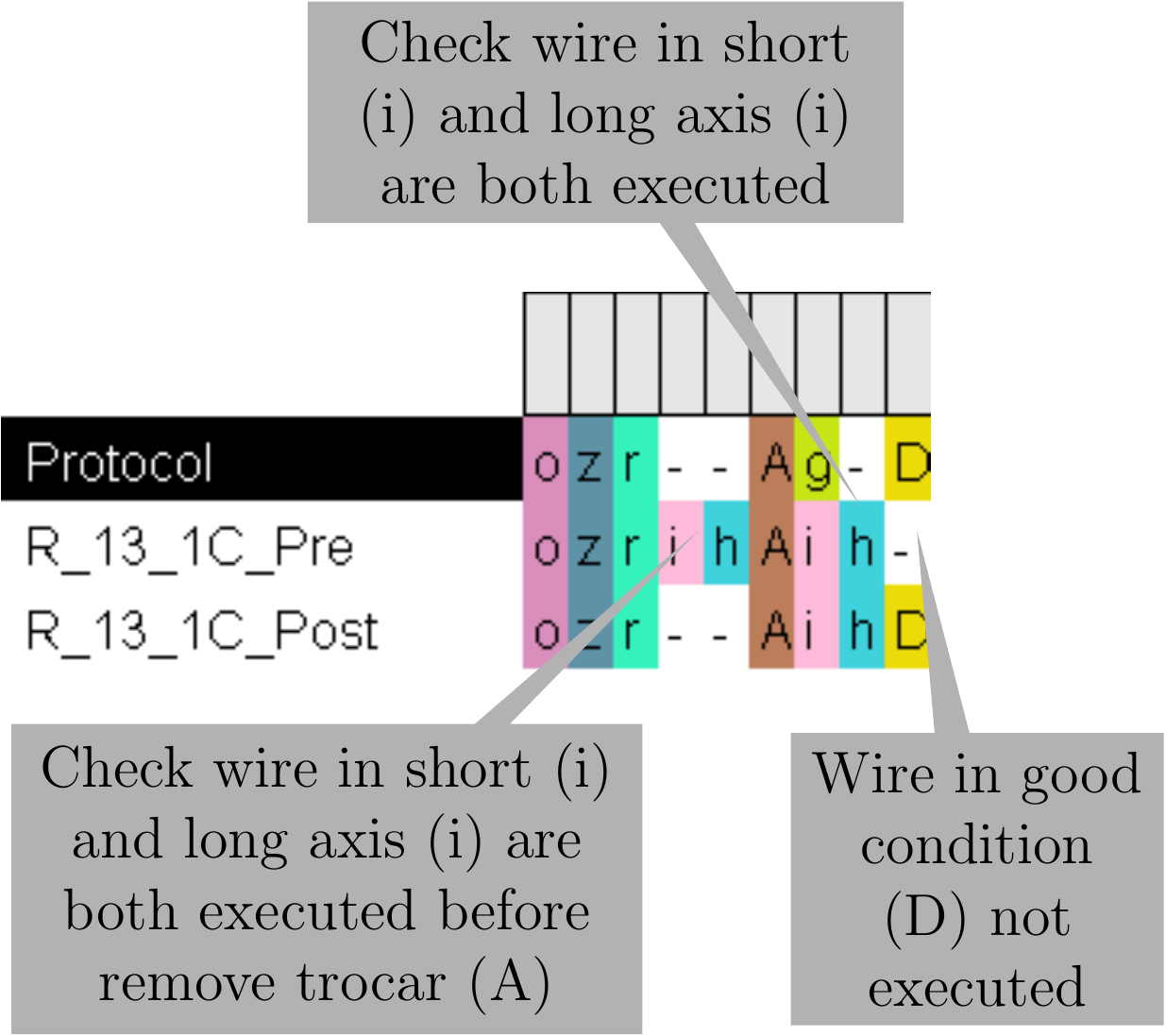}
\caption{Install Guidewire}
\label{fig:R_13_1C_Installguidewire}
\end{subfigure}
\hfill
\begin{subfigure}[t]{0.25\textwidth}
\includegraphics[width=0.99\linewidth]{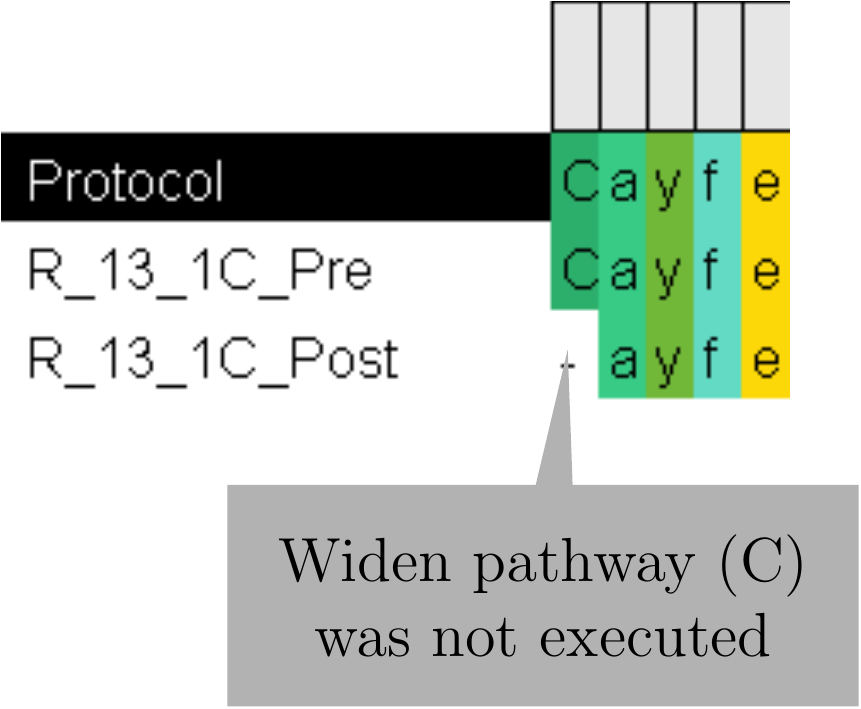}
\caption{Install catheter}
\label{fig:R_13_1C_InstallCatheter}
\end{subfigure}
\caption{Trace alignment of the sublogs pertaining to locate structures, install guidewire and install catheter stages by the student \texttt{\detokenize{R_13_1C}}.}
\label{fig:R_13_1CStageLevelAlignment}
\end{figure}

We can also analyze how a student performs in the individual stages using trace alignment on the sublogs for the various stages. This can help students analyze their behavior in segments on finer granular aspects. \figurename~\ref{fig:R_13_1C_TraceAlignmentSubLogOPUP}(a) and \figurename~\ref{fig:R_13_1C_TraceAlignmentSubLogOPUP}(b) depict the alignment of the traces pertaining to the operator and patient preparation and ultrasound configuration for the student \texttt{\detokenize{R_13_1C}}. As can been seen from the figure, \texttt{\detokenize{R_13_1C}} executes hand washing and get sterile clothes twice in the PRE round and the activity Prepare implements was executed last. While in the POST round, the only deviation is w.r.t the multiple execution of Prepare implements. \figurename~\ref{fig:R_13_1CStageLevelAlignment} depicts the alignment of traces pertaining to the student \texttt{\detokenize{R_13_1C}} for the stages locate structures, install guidewire, and install catheter. The prominent deviations are also highlighted in the figure.

\subsubsection{Time Analysis}
\begin{sidewaysfigure}[p]
\centering
{
\begin{subfigure}[t]{0.99\textwidth}
\includegraphics[width=0.99\linewidth]{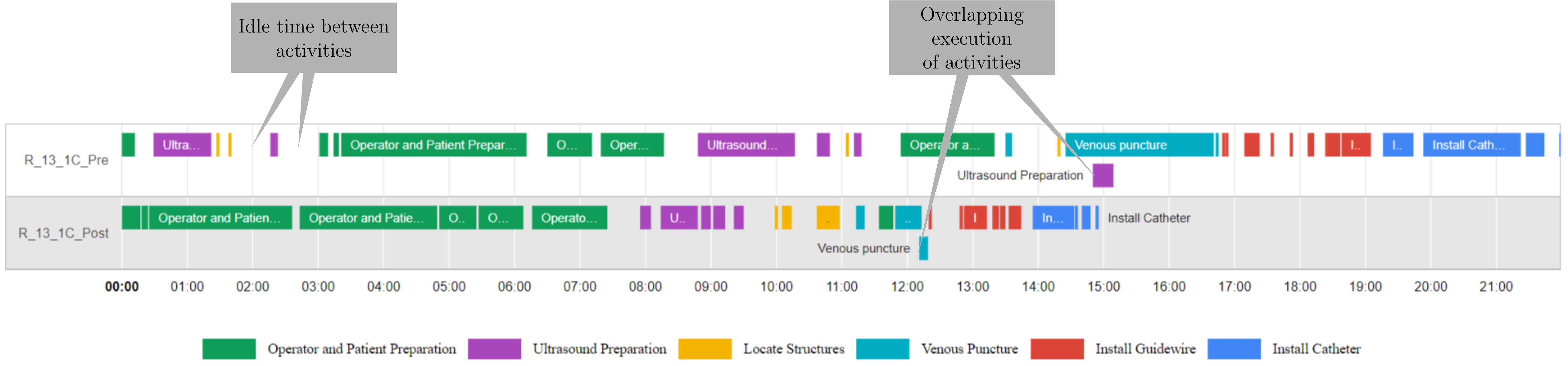}
\caption{Temporal view of the processing times for various stages for the student \texttt{\detokenize{R_13_1C}}.}
\label{fig:R_13_1C_GTL}
\end{subfigure}
}
\vfill
\vspace{1cm}
{
\begin{subfigure}[t]{0.32\textwidth}
\centering
\includegraphics[width=0.99\linewidth]{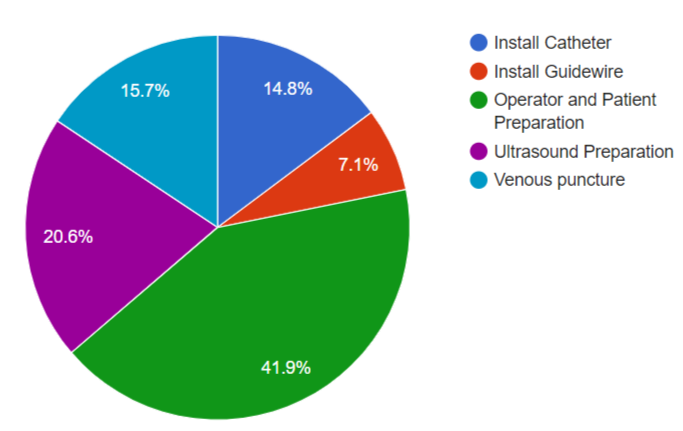}
\caption{Pre}
\label{fig:R_13_1C_PreStagePie}
\end{subfigure}
\hfill
\begin{subfigure}[t]{0.32\textwidth}
\centering
\includegraphics[width=0.99\linewidth]{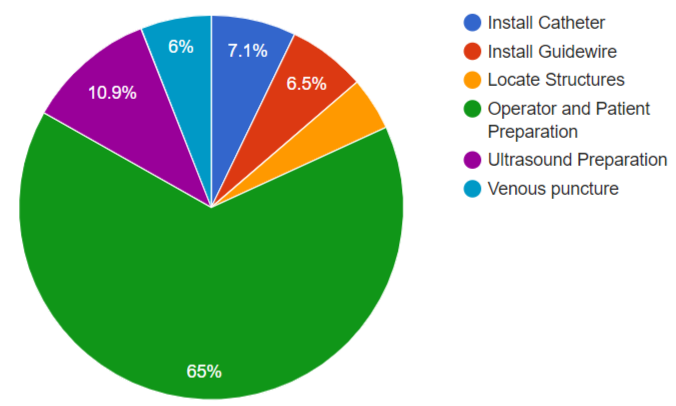}
\caption{Post}
\label{fig:R_13_1C_PostStagePie}
\end{subfigure}
\hfill
\begin{subfigure}[t]{0.32\textwidth}
\centering
\includegraphics[width=0.99\linewidth]{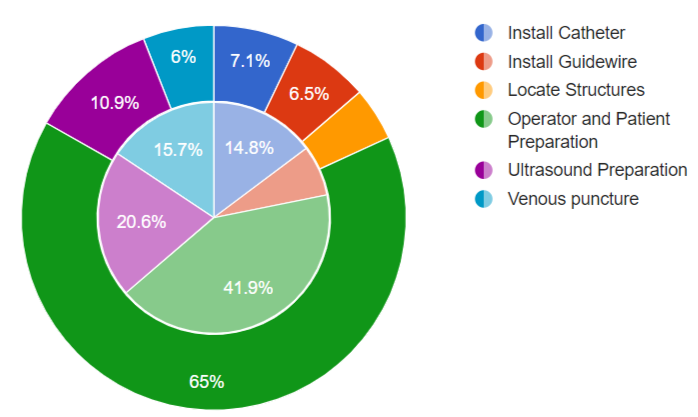}
\caption{Difference}
\label{fig:R_13_1C_DiffStagePie}
\end{subfigure}
}
\caption{Processing times spent in different stages by student \texttt{\detokenize{R_13_1C}}. }
\label{fig:R_13_1C_ProcessingTimeVis}
\end{sidewaysfigure}

\figurename~\ref{fig:R_13_1C_ProcessingTimeVis}(a) depicts the temporal view of the various stages performed by the student \texttt{\detokenize{R_13_1C}} during both the PRE and POST rounds. From the figure, we can see that the overall time taken by the student in the POST round is less than that of the PRE round. It is important to note that there are instances of overlapping activity executions in both the PRE and POST rounds. In the PRE round, it corresponds to events $35-38$, comprising of activities, Puncture-start, Ultrasound configuration-start, Ultrasound configuration-complete, and Puncture-complete, {\it belonging to two different stages}. {\it It needs to be validated by the domain experts whether such a behavior can be allowed or not}. Similarly, in the POST round, we observe overlapping activity executions in the events $35-38$, comprising of activities, Puncture-start, Blood return-start, Puncture-complete, Blood return-complete, all belonging to the Venous Puncture stage.

\figurename~\ref{fig:R_13_1C_ProcessingTimeVis}(b) and \figurename~\ref{fig:R_13_1C_ProcessingTimeVis}(c) depict the percentage of time spent by the student executing activities of the different stages during the PRE and POST rounds. Overall, \texttt{\detokenize{R_13_1C}} spent $1319$ secs and $894$ secs  doing the procedure in the PRE and POST rounds respectively. The student performed the procedure in $32\%$ less time in the POST round when compared to the PRE round. The student spent $907$ secs $(68.7\%)$ and $672$ secs $(75.2\%)$ (in the PRE and POST rounds respectively) performing some activity and the rest of the time has been idle. \figurename~\ref{fig:R_13_1C_ProcessingTimeVis}(d) provides a comparison on the amounts of time spent in different stages for the PRE and POST rounds. The inner circle corresponds to the PRE round. From the figure, we can see that in the PRE round $41.9\%$ of the total processing time was spent in the operator and patient preparation stage while in the POST round, it was $65\%$. Also, the amount of time spent in locate structures in the PRE round is $0$ (although the event log records activities from that stage, the VIDEOSTART and VIDEOEND time recorded is the same). Except for these two stages, the student spends lesser time in other stages in the POST round when compared with the PRE round. {\it If we have benchmark (protocol) expectations of time to be spent in different stages, we can do such analysis to identity the areas where the student either over performs or under performs and take corrective actions.}

In this fashion, one can analyze the execution traces of different students separately for their control-flow and time analysis. \appendixname~\ref{sec:appendtracealign} illustrates the trace alignment of different students w.r.t the protocol trace for both the PRE and POST rounds.

\subsection{Instructor Perspective}
In the previous section, we have looked at techniques that enable each individual student to analyze their performance. In this section, we consider the perspective (role) of the instructor and discuss approaches that can assist the instructors in analyzing their students performance. While the instructors too can use the analysis for individual students presented in the previous section to learn more about how each student performed, here we focus on {\it between student} analysis. Some of the questions that an instructor might be interested in gaining insights on include:
\begin{itemize}
\item whether there are students who struggle with the procedure (e.g., deviate a lot from the protocol, take longer times than usual, etc.)?
\item what sort of common mistakes/error do students generally do, where and how do they manifest?
\item are there any common patterns of behavior that students follow?
\item is there an improvement in students performance in the POST round when compared to the PRE round?
\end{itemize}

The learning from these insights can help the instructors adapt their teaching/learning methodology to ensure students are better skilled at executing these critical procedures.

\begin{figure}[!htb]
\centering
\includegraphics[width=0.8\linewidth]{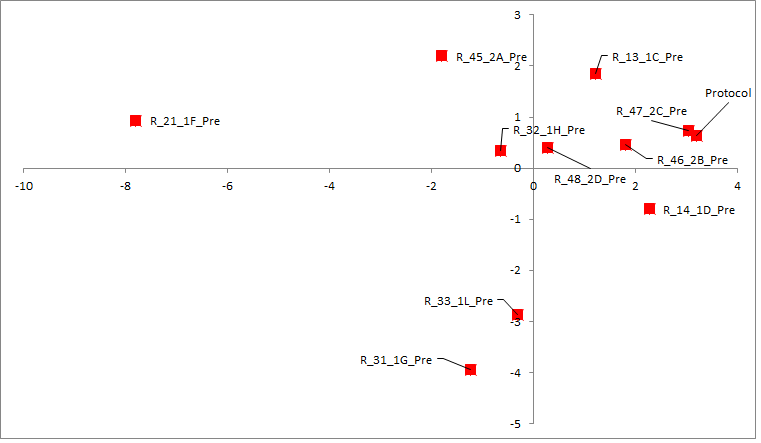}
\caption{Scatter plot of the principal components for the protocol and PRE round cases.}
\label{fig:pceigenpre}
\end{figure}

\figurename~\ref{fig:pceigenpre} depicts the scatter plot of the two principal components obtained using the Principal Component Analysis (PCA) \cite{pca} on the protocol trace and the PRE round cases of the 10 students. The protocol trace along with the 10 PRE round cases are first transformed into an {\it trace $\times$ activity frequency} matrix where each cell $ij$ of the matrix corresponds to the frequency of activity $j$ in trace $i$. PCA was applied on this {\it trace $\times$ activity frequency} matrix and the data is projected onto the top two principal components. {\it The scatter plots depict the proximity of traces with respect to each other and the protocol trace}. For example, from  \figurename~\ref{fig:pceigenpre}, we can see that \texttt{\detokenize{R_21_1F}} is the farthest from the protocol trace, implying that \texttt{\detokenize{R_21_1F}} has a lot of deviations from the protocol trace\footnote{Note that we have considered the activity frequency matrix for generating PCA. This implies that the proximity of traces in scatter plot is significant only with respect to their similarity considering the activities and their frequency. This does not take into account any ordering/dependenciees between the activities. Farther traces signify that they differ in the activities present and/or their frequencies. If we want to identify traces that are closer to the protocol trace even considering the ordering dependencies, we can use other representations of the event log that capture the ordering relations such as {\it trace $\times$ causal footprint} matrix where the columns correspond to ordering relations like {\it follows, precedes} (e.g., `a' follows `b')}, which is evident from the trace alignment of this case in \figurename~\ref{fig:R_21_1FPreTraceAlignment}. Similarly, the cases  \texttt{\detokenize{R_31_1G}} and \texttt{\detokenize{R_33_1L}} are more deviant from the protocol trace. On the other hand, the trace \texttt{\detokenize{R_47_2C}}, which is closest to the protocol trace, is expected to have less deviations (refer to \appendixname~\ref{sec:appendtracealign} for the trace alignments of these).

\begin{figure}[!htb]
\centering
\includegraphics[width=0.8\linewidth]{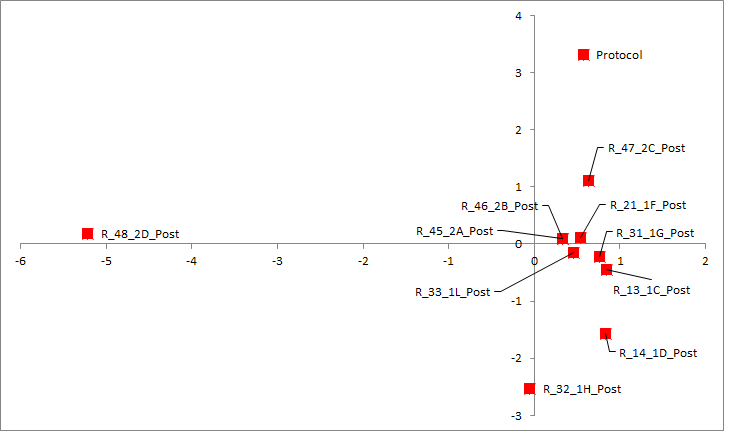}
\caption{Scatter plot of the principal components for the protocol and POST round cases.}
\label{fig:pceigenpost}
\end{figure}

\figurename~\ref{fig:pceigenpost} depicts the scatter plot of the two principal components on the protocol trace and the POST round cases of the 10 students using the {\it trace $\times$ activity frequency}.  While the traces that were more deviant (farther) in the PRE round such as \texttt{\detokenize{R_21_1F}} and \texttt{\detokenize{R_31_1G}} are closer to the protocol trace in the POST round, new cases have emerged to be more deviant. Traces pertaining to \texttt{\detokenize{R_48_2D}}, \texttt{\detokenize{R_32_1H}}, and \texttt{\detokenize{R_14_1D}} are farther to the protocol trace in the POST round implying they are more deviant (refer to \appendixname~\ref{sec:appendtracealign} for the trace alignments of these).

\subsubsection{Control-flow Analysis}
\begin{figure}[!htb]
\centering
\includegraphics[width=0.75\linewidth]{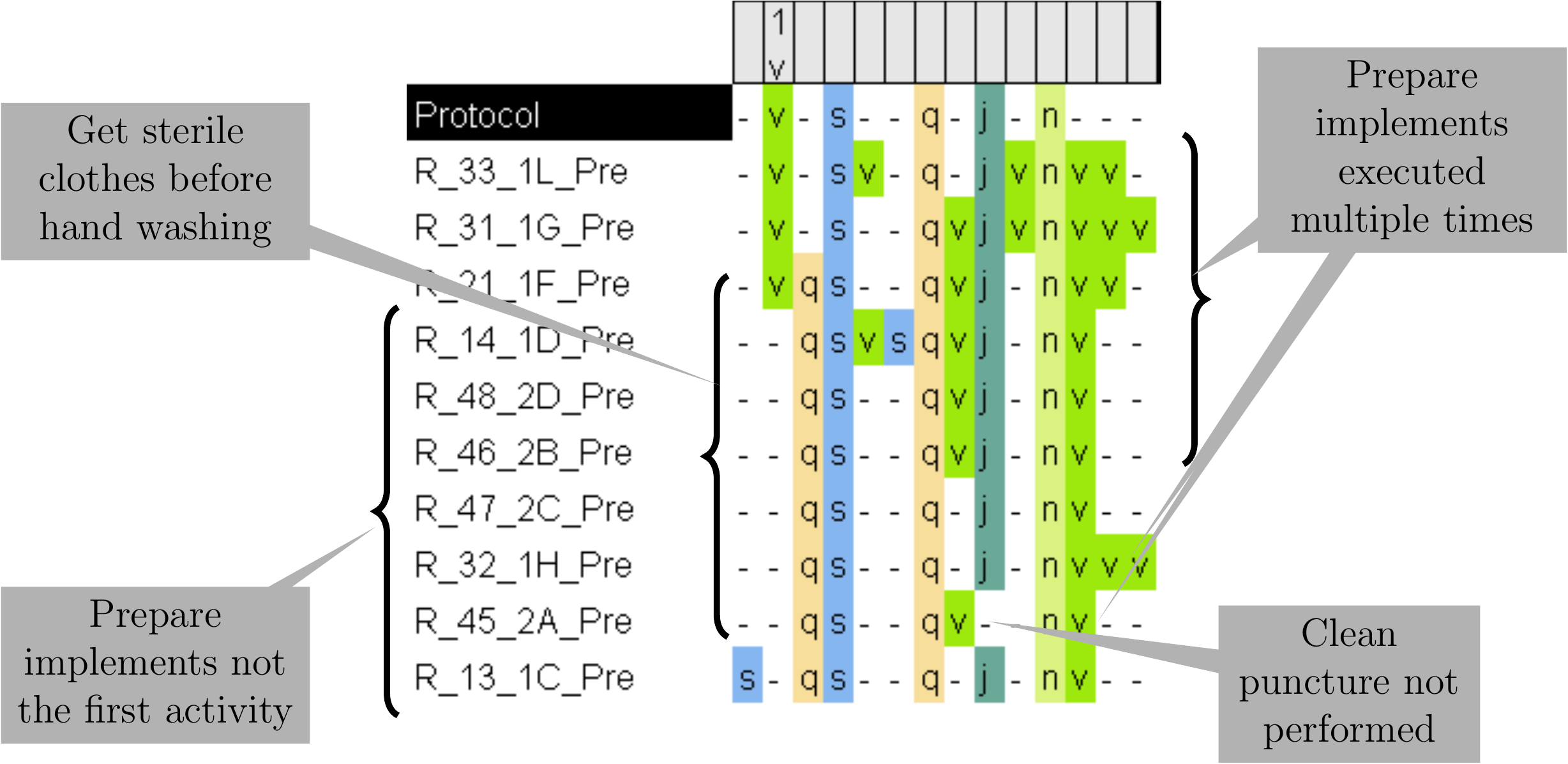}
\caption{Trace alignment of Operator and Patient Preparation sublog of the PRE round.}
\label{fig:preoperatortracealignment_allresources}
\end{figure}

~\newline
\textbf{Trace Alignment}\\
We first discuss the results of alignment of traces at stage level. \figurename~\ref{fig:preoperatortracealignment_allresources} depicts the trace alignment of the operator and patient preparation stage. the following deviations are observed from the alignment
\begin{itemize}
\item in 7 out of 10 cases, Get sterile clothes has been performed before Hand washing
\item in 7 out of 10 cases, Prepare implements is not the first activity
\item the student \texttt{\detokenize{R_45_2A}} didn't perform the activity Clean puncture area
\item except for \texttt{\detokenize{R_13_1C}} and \texttt{\detokenize{R_47_2C}}, the activity Prepare implements has been executed more than once
\item except for \texttt{\detokenize{R_31_1G}} and \texttt{\detokenize{R_33_1L}}, the activity Get sterile clothes has been executed twice
\item the activity Hand washing has been performed twice by \texttt{\detokenize{R_13_1C}} and \texttt{\detokenize{R_14_1D}}
\end{itemize}

\figurename~\ref{fig:preultrasoundtracealignment_allresources} depicts the alignment of traces in the sublog pertaining to the ultrasound preparation stage during the PRE round. The following deviations are observed in the traces:
\begin{figure}[!htb]
\centering
\includegraphics[width=0.75\linewidth]{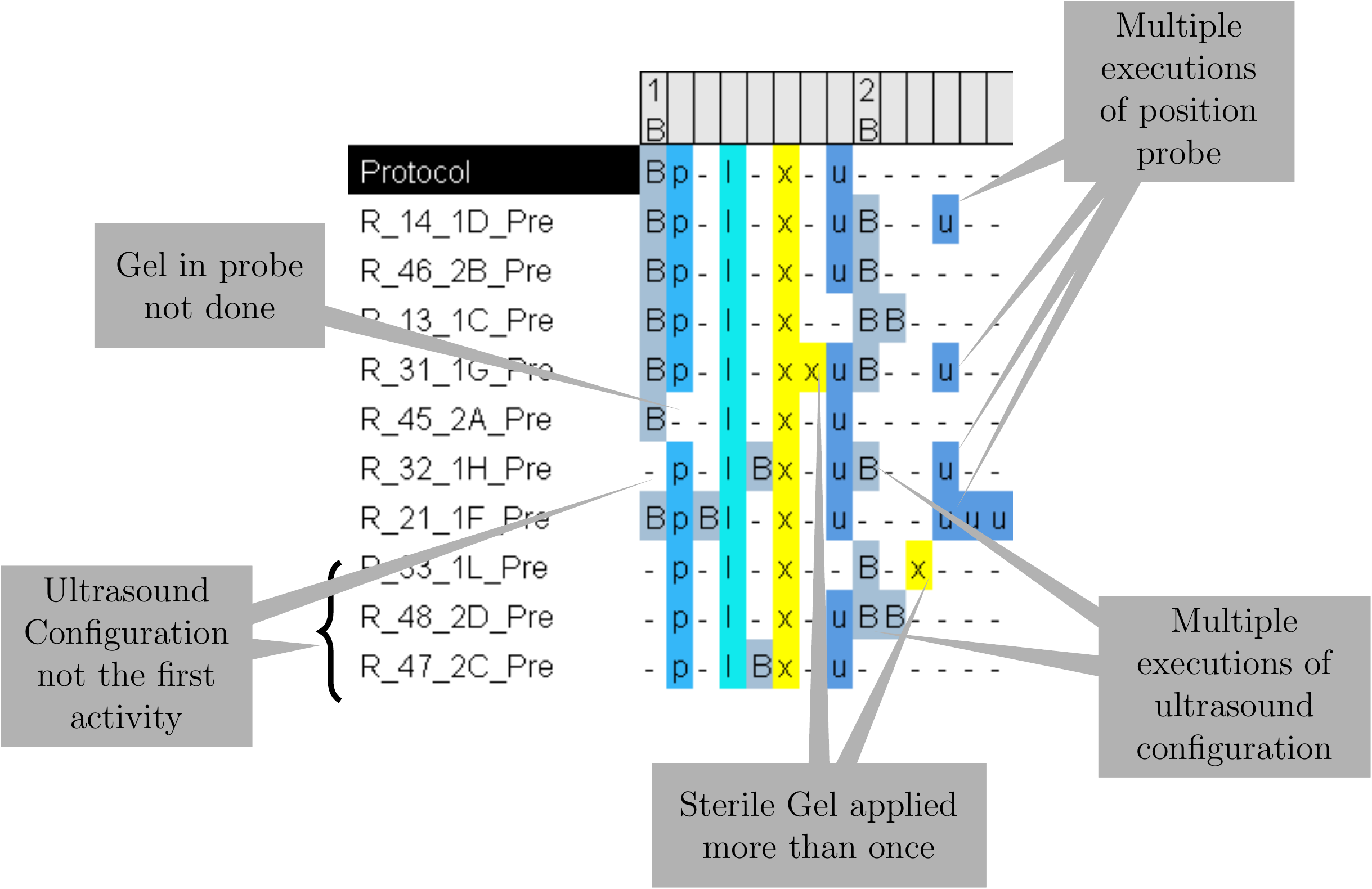}
\caption{Trace alignment of Ultrasound Preparation sublog of the PRE round.}
\label{fig:preultrasoundtracealignment_allresources}
\end{figure}
\begin{itemize}
\item in 4 out of 10 cases, ultrasound configuration is not the first activity
\item the activity Gel in probe has not been performed by student \texttt{\detokenize{R_45_2A}}
\item the activity Position probe has not been done by two students \texttt{\detokenize{R_13_1C}} and \texttt{\detokenize{R_33_1L}}
\item the activity Put sterile gel has been executed twice by two students \texttt{\detokenize{R_31_1G}} and \texttt{\detokenize{R_33_1L}}
\item in 4 out of 10 cases, the activity Position probe has been executed twice
\item except for \texttt{\detokenize{R_45_2A}}, \texttt{\detokenize{R_33_1L}}, and \texttt{\detokenize{R_47_2C}}, the activity ultrasound configuration has been executed more than once
\end{itemize}
\begin{figure}[!htb]
\begin{subfigure}{0.5\textwidth}
\centering
\includegraphics[width=0.99\linewidth]{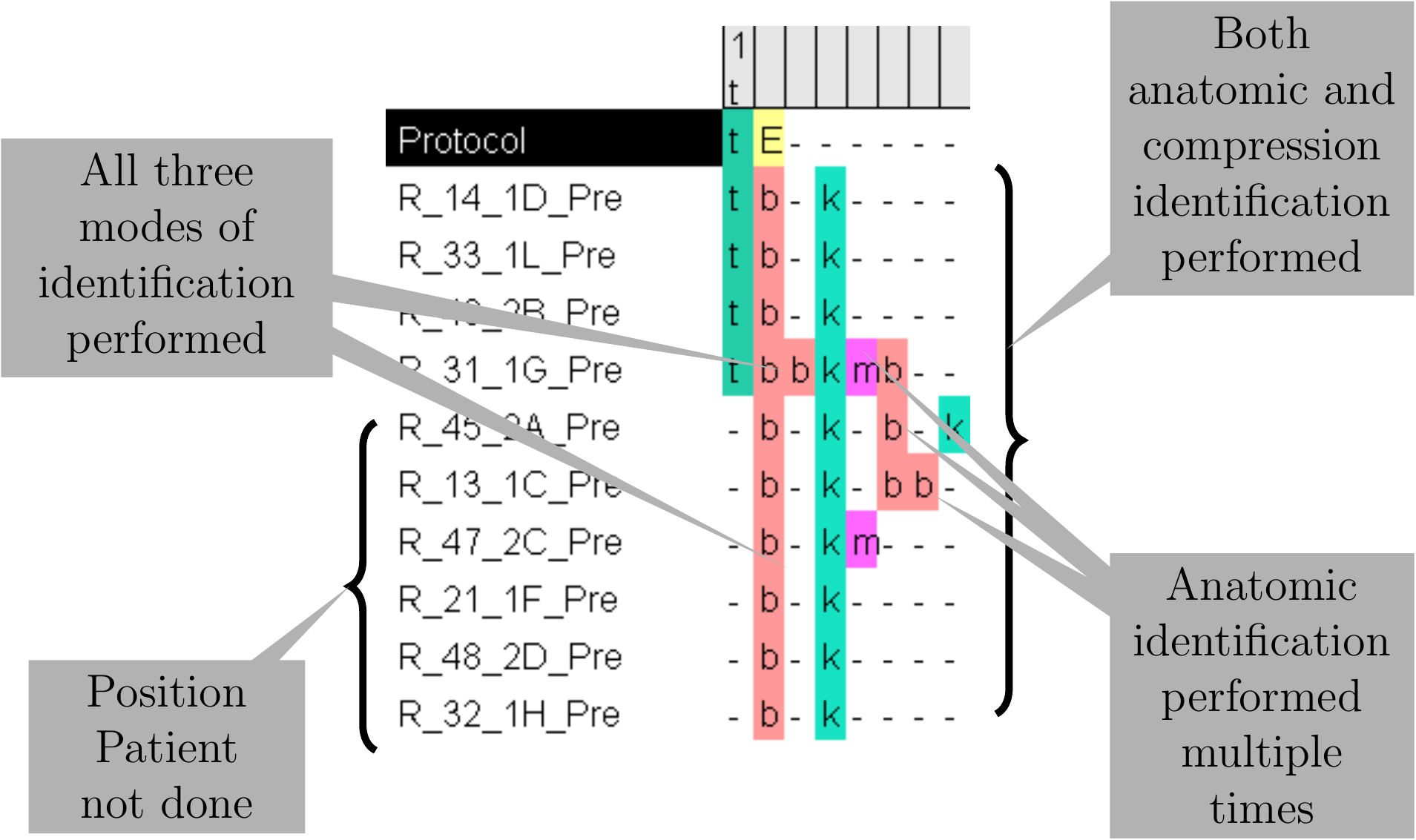}
\caption{Locate Structures}
\label{fig:prelocatetracealignment_allresources}
\end{subfigure}
\hfill
\begin{subfigure}{0.35\textwidth}
\centering
\includegraphics[width=0.99\linewidth]{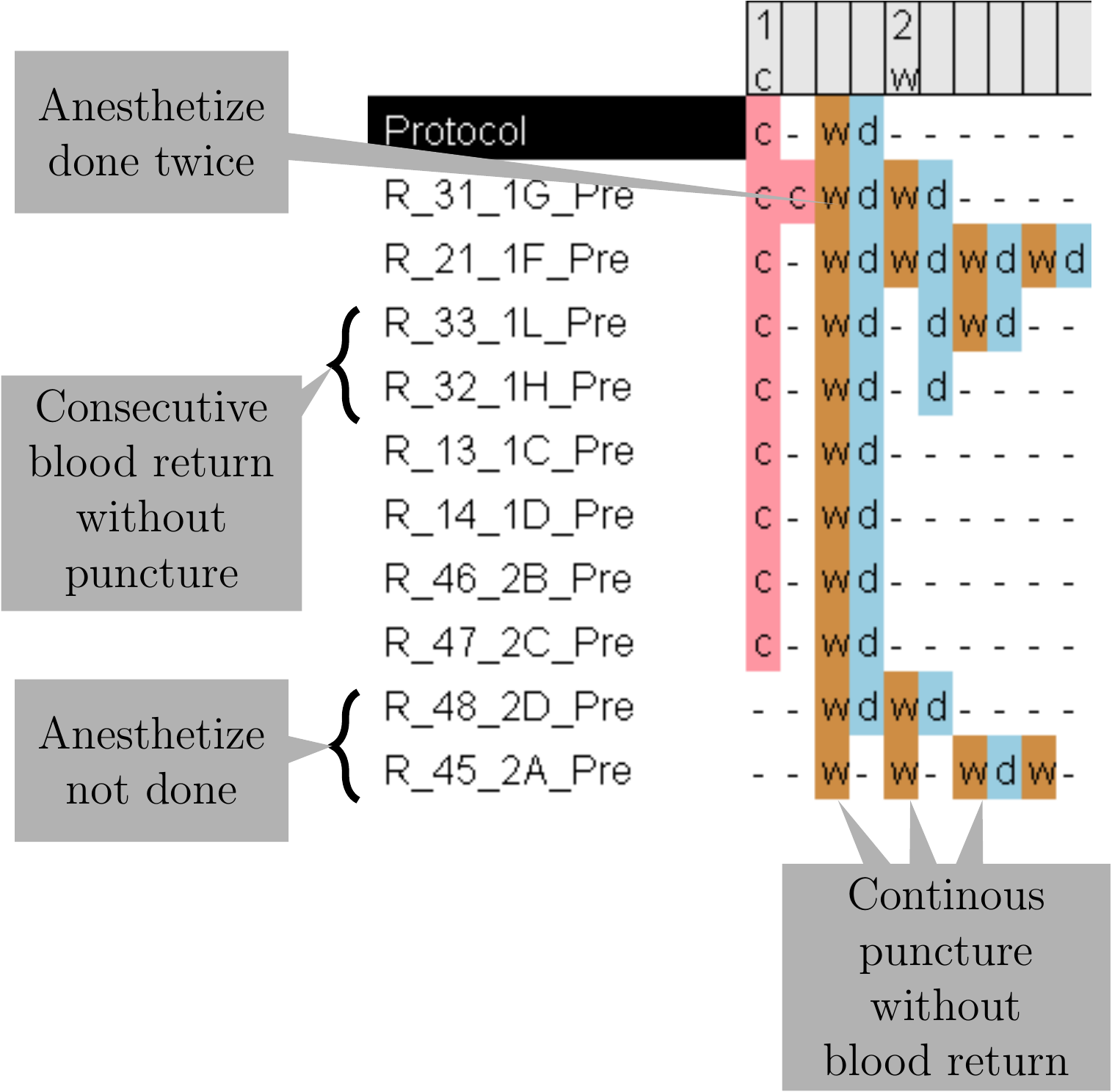}
\caption{Venous Puncture}
\label{fig:prevenoustracealignment_allresources}
\end{subfigure}
\caption{Trace alignment of locate structures and venous puncture sublogs of the PRE round.}
\label{fig:prelocatevenoustracealignment_allresources}
\end{figure}
\figurename~\ref{fig:prelocatetracealignment_allresources} depicts the alignment of the locate structures sublog of the PRE round. The following deviations are observed in this stage: (i) in 6 out of 10 cases, Position patient has not been performed, (ii) in all the cases both anatomic identification and compression identification has been performed, (iii) students \texttt{\detokenize{R_31_1G}} and \texttt{\detokenize{R_47_2C}} execute all three identification modes, and (iv) anatomic identification has been executed more than once by \texttt{\detokenize{R_31_1G}}, \texttt{\detokenize{R_45_2A}}, and \texttt{\detokenize{R_13_1C}} while compression identification has been executed twice by \texttt{\detokenize{R_45_2A}}.

\figurename~\ref{fig:prevenoustracealignment_allresources} depicts the alignment of the venous puncture sublog of the PRE round. The
following deviations are observed in this stage: (i) student \texttt{\detokenize{R_31_1G}} has performed Anesthetize twice, (ii) students
\texttt{\detokenize{R_45_2A}} and \texttt{\detokenize{R_48_2D}} didn't perform Anesthetize, (iii) student \texttt{\detokenize{R_45_2A}} performed continuous Puncture without checking Blood return, and (iv) students \texttt{\detokenize{R_32_1H}} and \texttt{\detokenize{R_33_1L}} perform two continuous blood return checks without a puncture. Note that the reference model allows a loop over Puncture and Blood return activities, i.e., loops over `wd' activities in the alignment. Although the protocol trace captured only one instance of `wd', while interpreting the alignment, we consider multiple executions of  `wd' to be compliant to the protocol.
\begin{figure}[!htb]
\begin{subfigure}[e]{0.64\textwidth}
\centering
\includegraphics[width=0.99\linewidth]{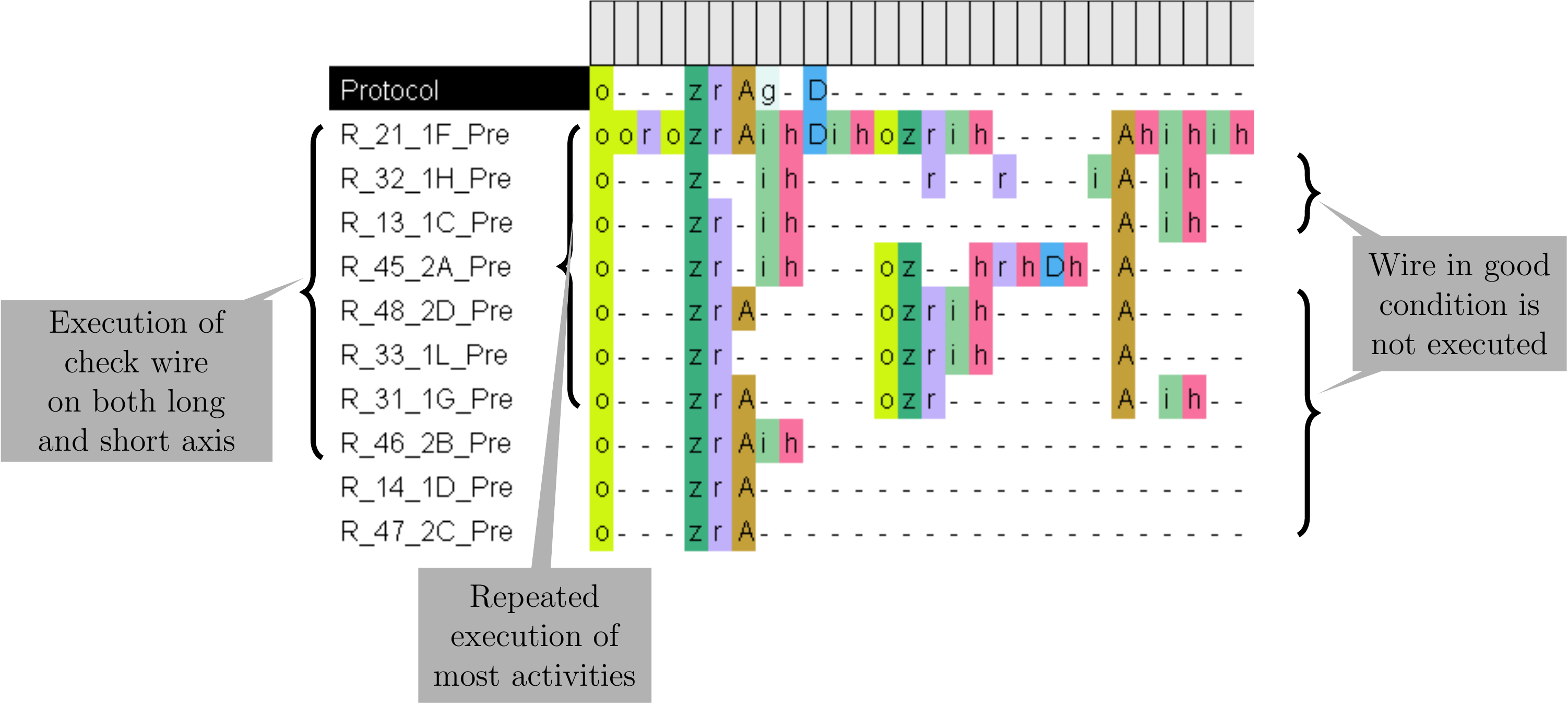}
\caption{Install Guidewire}
\label{fig:preguidewiretracealignment_allresources}
\end{subfigure}
\hfill
\begin{subfigure}[f]{0.35\textwidth}
\centering
\includegraphics[width=0.99\linewidth]{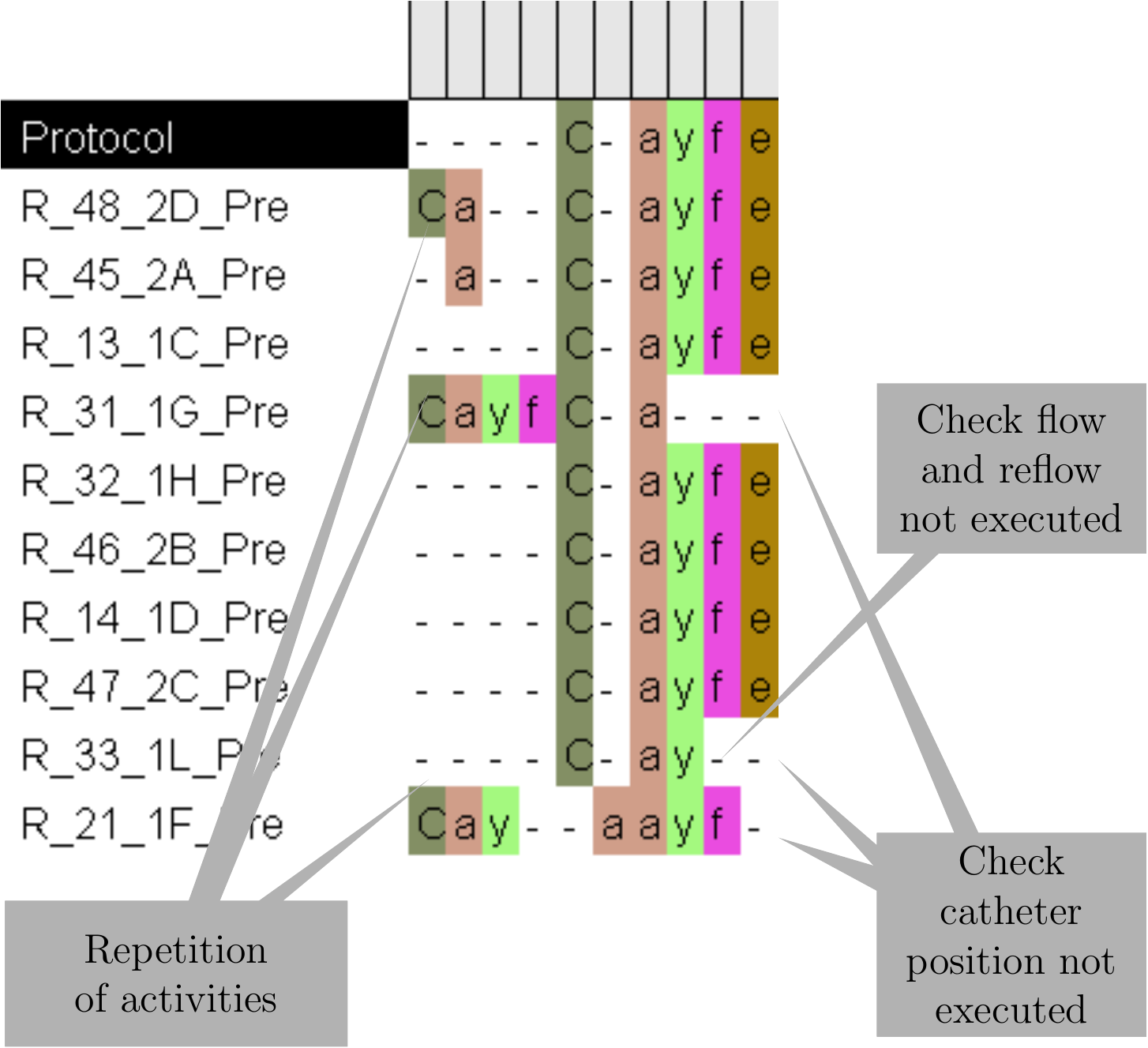}
\caption{Install Catheter}
\label{fig:precathetertracealignment_allresources}
\end{subfigure}
\caption{Trace alignment of install guidewire and install catheter sublogs of the PRE round.}
\label{fig:preguidewirecatheterTraceAlignment}
\end{figure}

\figurename~\ref{fig:preguidewiretracealignment_allresources} depicts the alignment of the install guidewire sublog of the PRE round. The following deviations are observed in this stage: (i) the activity `wire in good condition' was not executed in $7$ of the $10$ traces (ii) in $8$ of the $10$ traces, check wire on both long and short axis were executed and in the other two, neither of them was executed. \figurename~\ref{fig:precathetertracealignment_allresources} depicts the alignment of the install catheter sublog of the PRE round. We notice the following deviations: (i) the activity check catheter position was not executed in $3$ traces (ii) two traces had executed Widen pathway and advance catheter twice while one trace had executed advance catheter thrice, (iii) \texttt{\detokenize{R_45_2A}} executed advance catheter before widening the pathway, and (iv) the activity check flow and reflow was not executed by \texttt{\detokenize{R_33_1L}}.

\begin{sidewaysfigure}[p]
\centering
\includegraphics[width=0.99\textwidth]{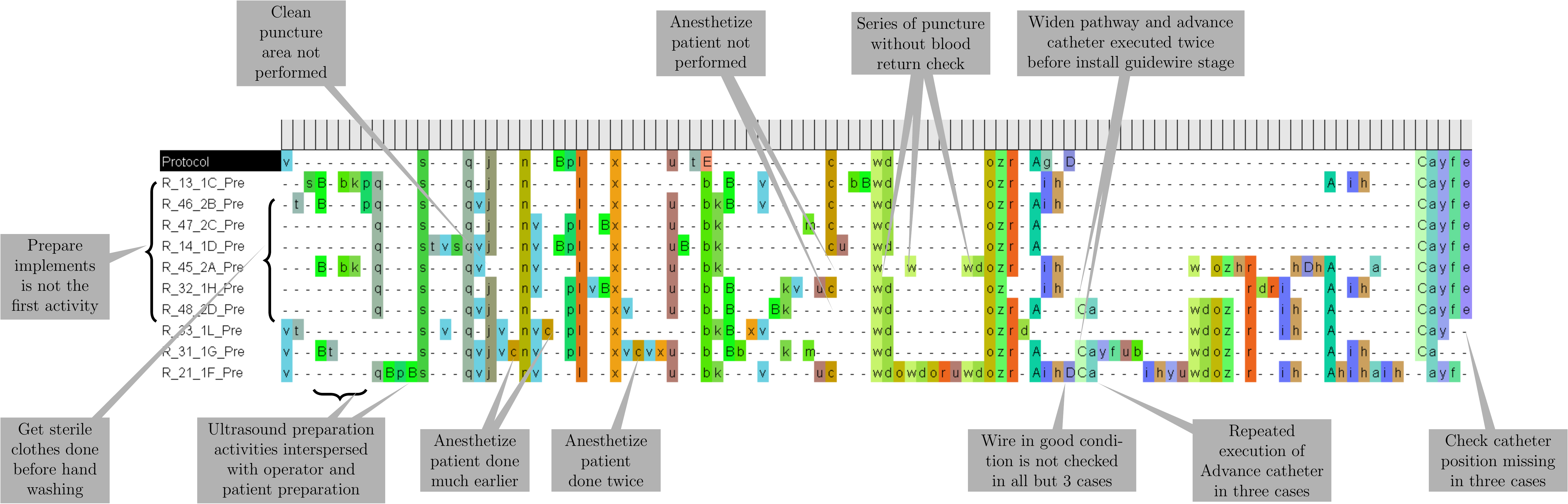}
\caption{Trace alignment of the protocol and student cases of the PRE round.}
\label{fig:preallresourcetracealignment}
\end{sidewaysfigure}

\figurename~\ref{fig:preallresourcetracealignment} depicts the trace alignment obtained by aligning the 10 student traces of the PRE round with the protocol trace. A common deviation uncovered by this alignment is the interspersed execution of activities belonging to different stages.
\begin{sidewaysfigure}[p]
\begin{subfigure}[b]{0.33\textwidth}
\centering
\includegraphics[width=0.99\linewidth]{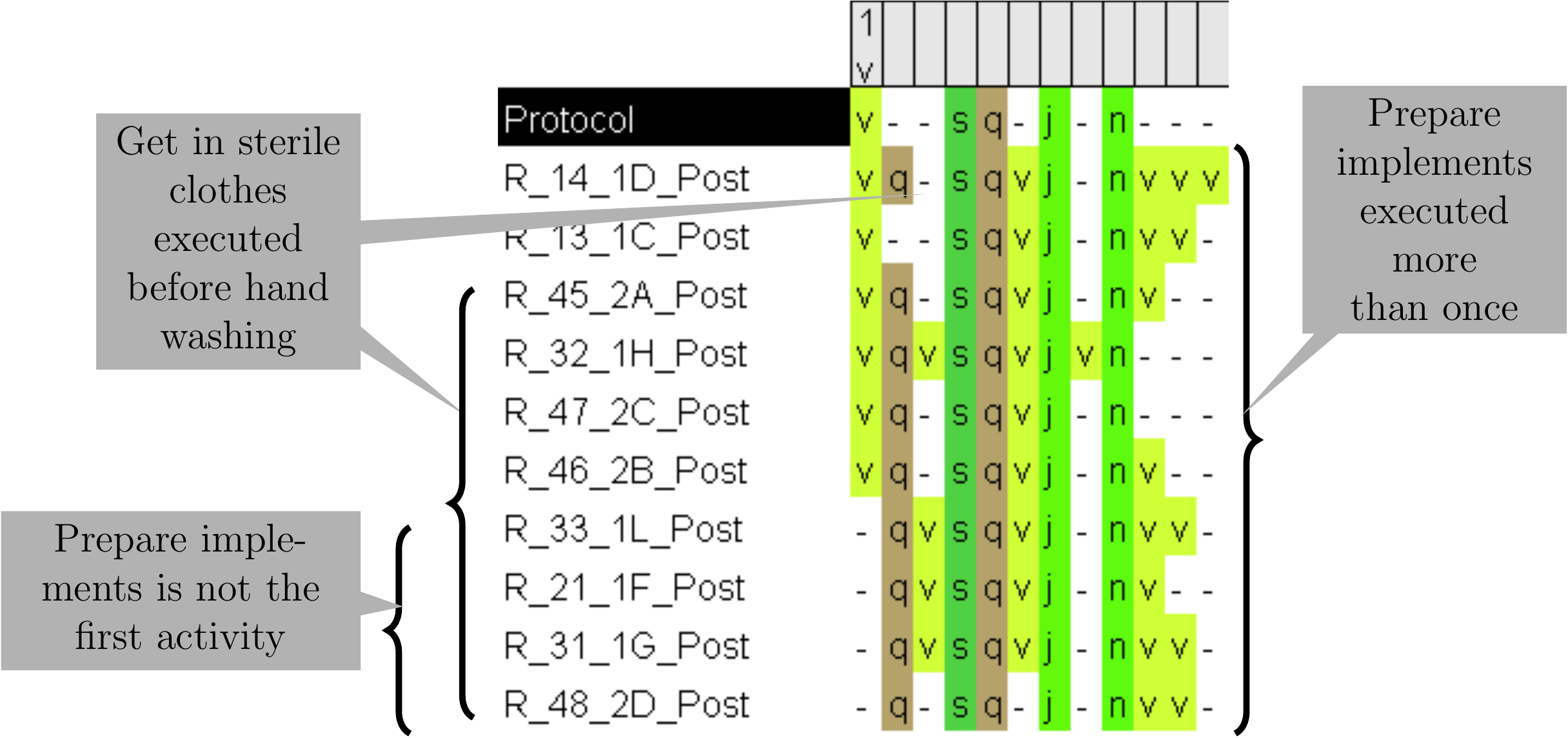}
\caption{Operator and Patient Preparation}
\label{fig:postoperatortracealignment_allresources}
\end{subfigure}
%
\begin{subfigure}[b]{0.33\textwidth}
\centering
\includegraphics[width=0.99\linewidth]{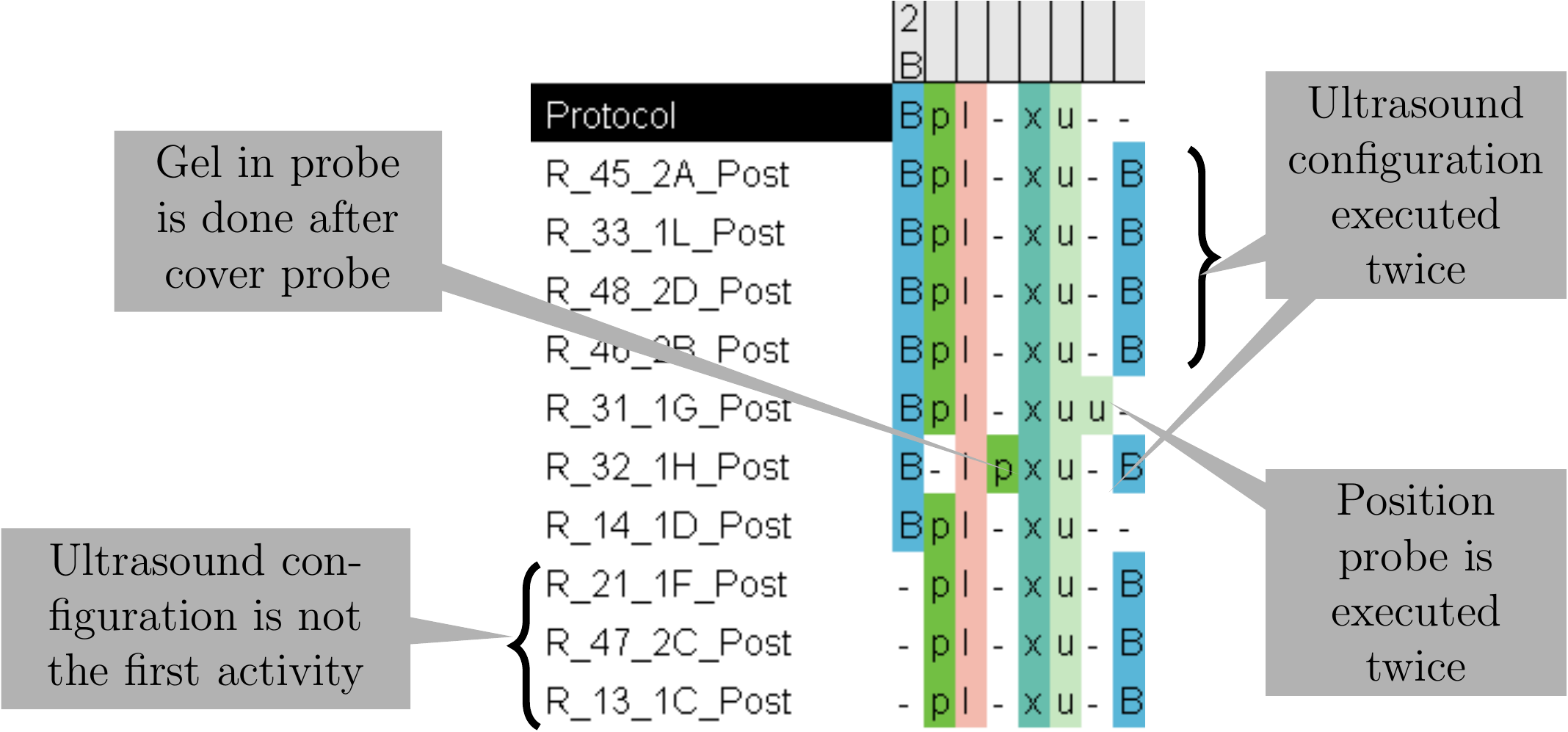}
\caption{Ultrasound Preparation}
\label{fig:postultrasoundtracealignment_allresources}
\end{subfigure}
%
\begin{subfigure}[b]{0.33\textwidth}
\centering
\includegraphics[width=0.99\linewidth]{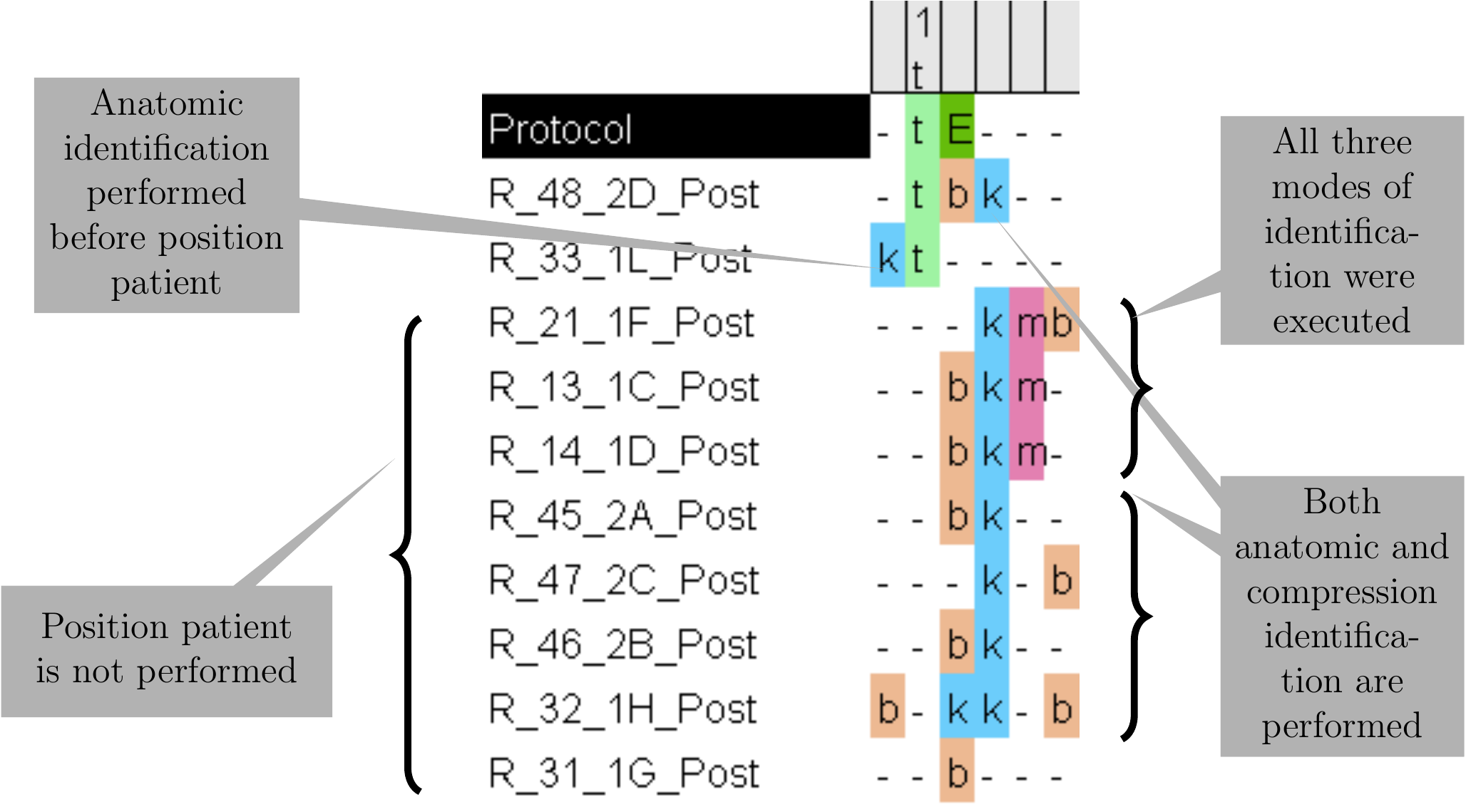}
\caption{Locate Structures}
\label{fig:postlocatetracealignment_allresources}
\end{subfigure}
%
\begin{subfigure}[b]{0.2\textwidth}
\centering
\includegraphics[width=0.99\linewidth]{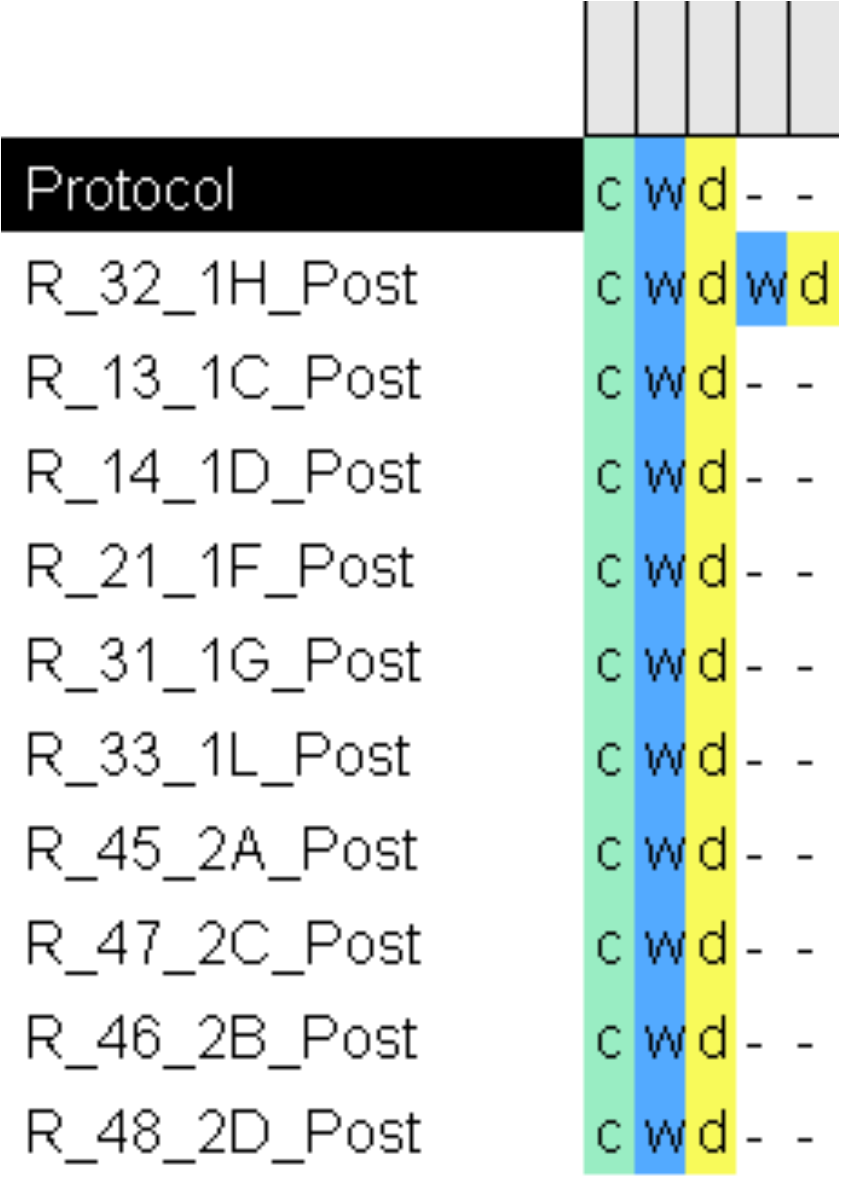}
\caption{Venous Puncture}
\label{fig:postvenoustracealignment_allresources}
\end{subfigure}
%
\begin{subfigure}[b]{0.33\textwidth}
\centering
\includegraphics[width=0.99\linewidth]{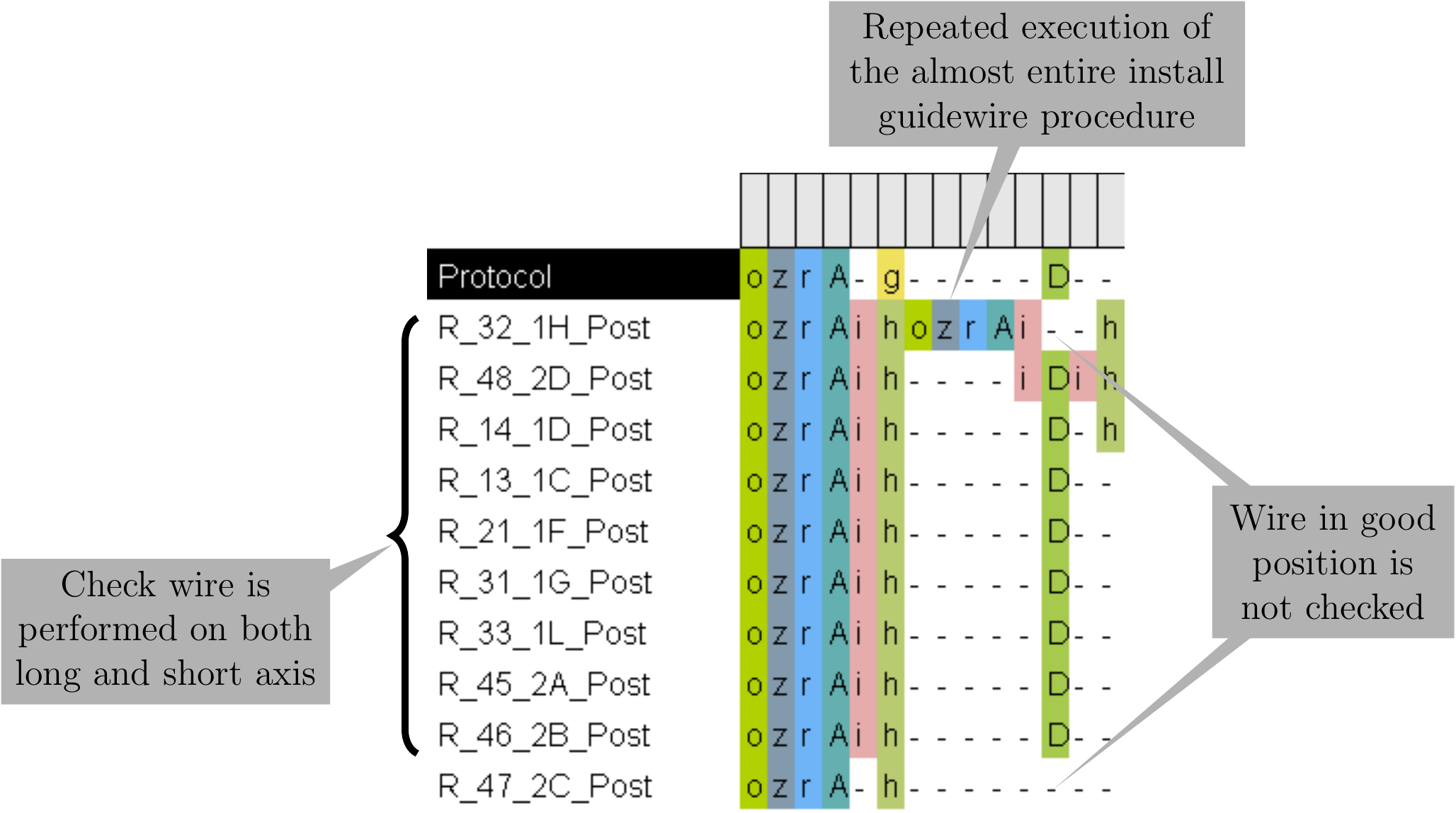}
\caption{Install Guidewire}
\label{fig:postguidewiretracealignment_allresources}
\end{subfigure}
%
\begin{subfigure}[b]{0.33\textwidth}
\centering
\includegraphics[width=0.99\linewidth]{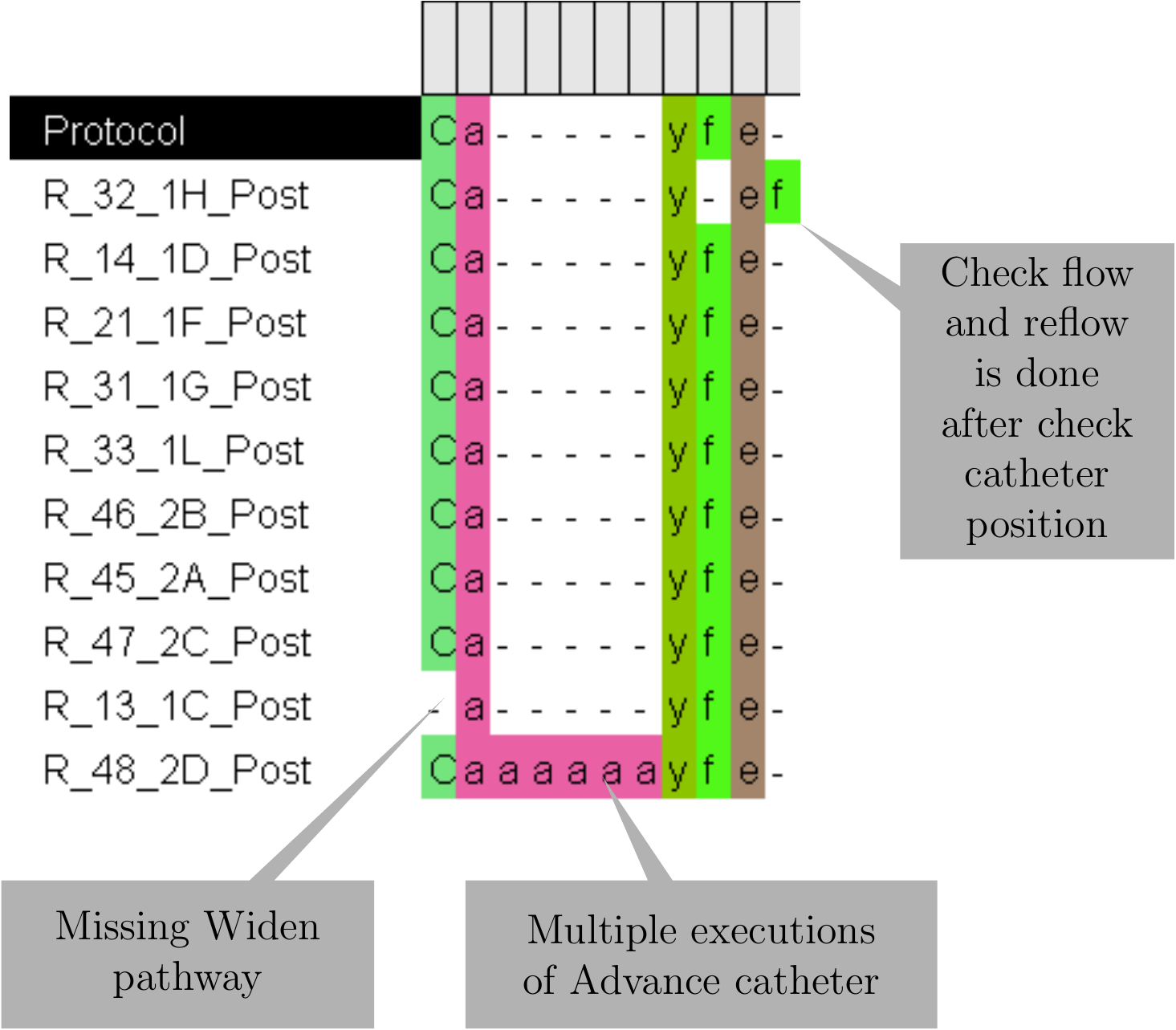}
\caption{Install Catheter}
\label{fig:postcathetertracealignment_allresources}
\end{subfigure}
\caption{Trace alignment of the different stages using the sublogs of the POST round}
\label{fig:postsublogalignment}
\end{sidewaysfigure}

\begin{sidewaysfigure}[p]
\centering
\includegraphics[width=0.99\textwidth]{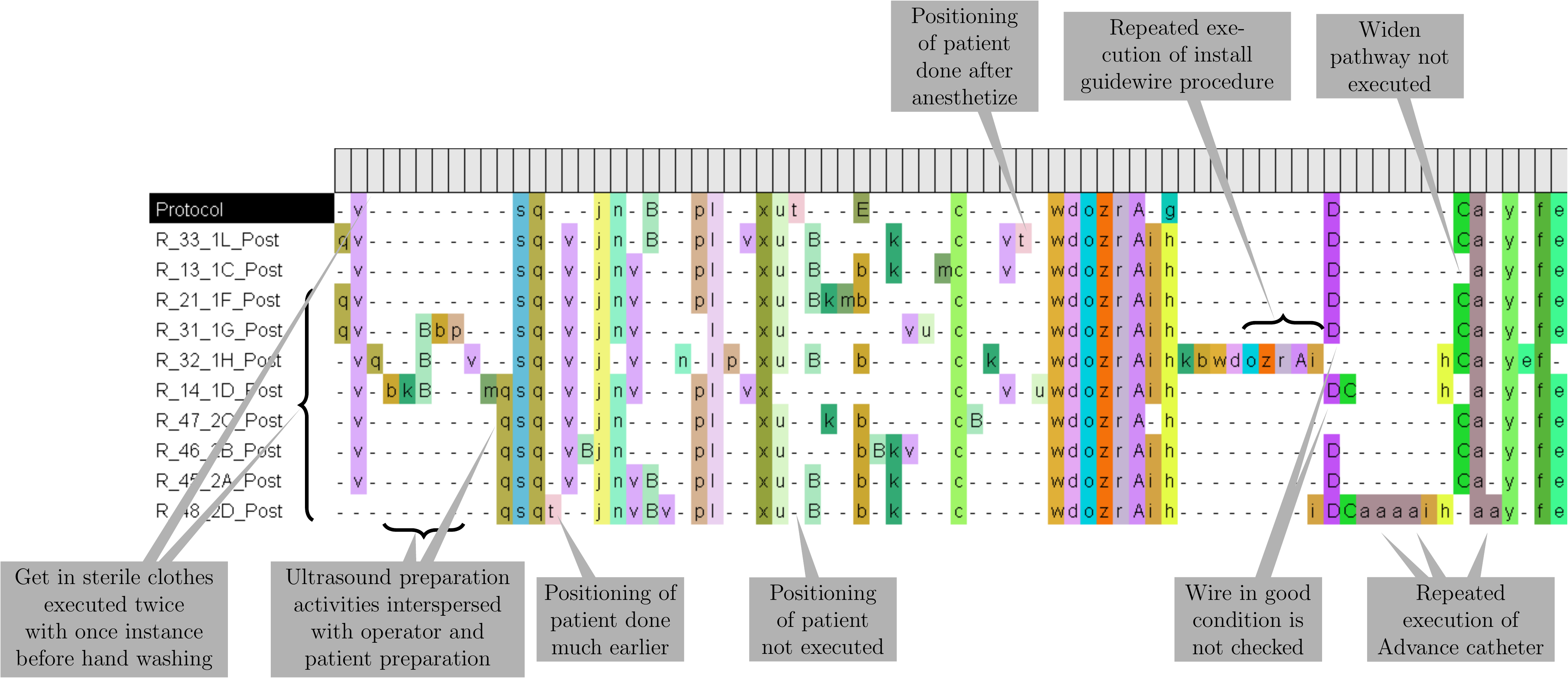}
\caption{Trace alignment of the complete procedure followed in the POST round.}
\label{fig:posttracealignment}
\end{sidewaysfigure}

\figurename~\ref{fig:postsublogalignment} depicts the alignment of the traces for the different stages in the POST round. The deviations are highlighted in the figure. As can be seen, the number of deviations is significantly less when compared to the PRE round. We see a lot of conserved regions in the POST round. \figurename~\ref{fig:posttracealignment} depicts the alignment of the complete traces followed in the POST round. Here again, we see fewer deviations when compared to the PRE round.

\textbf{Declarative Modeling and Analysis}
\tablename~\ref{tab:declarepre} specifies $10$ DECLARE constraints of the CVC procedure and the outcome of conformance analysis on the PRE round cases. An `x' in a cell of the table indicates that the corresponding trace (column name) is non-complaint w.r.t the relation specified in that row. For example, the execution trace of student \texttt{\detokenize{R_31_1G}} does not satisfy 4 of the 10 constriants viz., (a) the exclusive choice 1 of 3 constraint on the different modes of identification, (b) the exactly 1 constraint on anesthetize activity, (c) the constraint that one has to check whether guidewire is in good condition before advancing the catheter, and (d) the constraint that whenever install guidewire activity is executed, it should be followed by a remove guidewire activity. From the table, we can see that the constraints in rows $3$, $8$, and $1$ are the most violated ones with the constraint $3$ being violated by all the cases.
\begin{table}[!htb]
\centering
\caption{Declarative constraints of CVC procedure and the compliance analysis of the PRE round cases.}
\label{tab:declarepre}
\begin{tabular}{|l|l|c|c|c|c|c|c|c|c|c|c|c|}
\hline
\hline
S.No & Relation  & ~\rotatebox[origin=c]{90}{~\texttt{\detokenize{R_13_1C_Pre}}~}~ & ~\rotatebox[origin=c]{90}{~\texttt{\detokenize{R_14_1D_Pre}}~}~ & ~\rotatebox[origin=c]{90}{~\texttt{\detokenize{R_21_1F_Pre}}~}~& ~\rotatebox[origin=c]{90}{~\texttt{\detokenize{R_31_1G_Pre}}~}~& ~\rotatebox[origin=c]{90}{~\texttt{\detokenize{R_32_1H_Pre}}~}~& ~\rotatebox[origin=c]{90}{~\texttt{\detokenize{R_33_1L_Pre}}~}~& ~\rotatebox[origin=c]{90}{~\texttt{\detokenize{R_45_2A_Pre}}~}~ &
~\rotatebox[origin=c]{90}{~\texttt{\detokenize{R_46_2B_Pre}}~}~ & ~\rotatebox[origin=c]{90}{~\texttt{\detokenize{R_47_2C_Pre}}~}~ &
~\rotatebox[origin=c]{90}{~\texttt{\detokenize{R_48_2D_Pre}}~}~ & ~\rotatebox[origin=c]{90}{~Total~}\\
\hline
\hline
1 & Precedence(Hand washing, Get sterile clothes) & &x&x&&x&&x&x&x&x & {\bf 7}\\
\hline
\multirow{3}{*}{2} & Precedence(Position patient, Anatomic identification) &\multirow{3}{*}{x}&\multirow{3}{*}{}&\multirow{3}{*}{x}&\multirow{3}{*}{}&\multirow{3}{*}{x}&\multirow{3}{*}{}&\multirow{3}{*}{x}&\multirow{3}{*}{}&\multirow{3}{*}{x}&\multirow{3}{*}{x} &\multirow{3}{*}{\bf 6}\\
 & Precedence(Position patient, Doppler identification) & & & & & & & & & & & \\
 & Precedence(Position patient, Compression identification) & & & & & & & & & & & \\
\hline
\multirow{2}{*}{3} & Exclusive Choice 1~of~3 (Anatomic identification, & \multirow{2}{*}{x}&\multirow{2}{*}{x}&\multirow{2}{*}{x}&\multirow{2}{*}{x}&\multirow{2}{*}{x}&\multirow{2}{*}{x}&\multirow{2}{*}{x}&\multirow{2}{*}{x}&\multirow{2}{*}{x}&\multirow{2}{*}{x} & \multirow{2}{*}{\bf 10}\\
& Doppler identification, Compression identification) & & & & & & & & & & &\\
\hline
4 & Exactly 1 (Anesthetize) & &&&x&&&x&&&x & {\bf 3}\\
\hline
5 & Precedence(Anesthetize,Puncture) &&&&&&&x&&&x & {\bf 2}\\
\hline
6 & Alternate Response(Puncture,  Blood return) &&&&&&&x&&& & {\bf 1}\\
\hline
7 & Precedence(Remove Trocar, Wire in good position) & x&&&&x&x&x&&& & {\bf 4}\\
\hline
8 & Precedence(Wire in good position, Advance catheter) & x&x&&x&x&x&&x&x&x & {\bf 8}\\
\hline
9 & Response(Remove guidewire, Check flow and reflow) &&&&&&x&&&& & {\bf 1}\\
\hline
10 & Response(Install guidewire,  Remove guidewire) &&&&x&&&&&& & {\bf 1} \\
\hline
\hline
& {\bf Total} & \bf{4}&\bf{3}&\bf{3}&\bf{4}&\bf{5}&\bf{4}&\bf{7}&\bf{3}&\bf{4}&\bf{6} & {\bf 43}\\
\hline
\hline
\end{tabular}
\end{table}

\tablename~\ref{tab:declarepost} specifies the same $10$ DECLARE constraints of the CVC procedure and the outcome of conformance analysis on the POST round cases. We could see from the table that the overall number of violations across all constraints has reduced to $28$ in the POST round when compared to $43$ in the PRE round. Several of the constraints ($4-7$, $9$ and $10$) are fully satisfied. However, it is to be noted that the violations have increased for constraints $1$ and $2$ and reduced by just $1$ for constraint $3$.
\begin{table}[!htb]
\centering
\caption{Declarative constraints of CVC procedure and the compliance analysis of the POST round cases.}
\label{tab:declarepost}
\begin{tabular}{|l|l|c|c|c|c|c|c|c|c|c|c|c|}
\hline
\hline
S.No & Relation  & ~\rotatebox[origin=c]{90}{~\texttt{\detokenize{R_13_1C_Post}}~}~ & ~\rotatebox[origin=c]{90}{~\texttt{\detokenize{R_14_1D_Post}}~}~ & ~\rotatebox[origin=c]{90}{~\texttt{\detokenize{R_21_1F_Post}}~}~& ~\rotatebox[origin=c]{90}{~\texttt{\detokenize{R_31_1G_Post}}~}~& ~\rotatebox[origin=c]{90}{~\texttt{\detokenize{R_32_1H_Post}}~}~& ~\rotatebox[origin=c]{90}{~\texttt{\detokenize{R_33_1L_Post}}~}~& ~\rotatebox[origin=c]{90}{~\texttt{\detokenize{R_45_2A_Post}}~}~ &
~\rotatebox[origin=c]{90}{~\texttt{\detokenize{R_46_2B_Post}}~}~ & ~\rotatebox[origin=c]{90}{~\texttt{\detokenize{R_47_2C_Post}}~}~ &
~\rotatebox[origin=c]{90}{~\texttt{\detokenize{R_48_2D_Post}}~}~ & ~\rotatebox[origin=c]{90}{~Total~}\\
\hline
\hline
1 & Precedence(Hand washing, Get sterile clothes) &&x&x&x&x&x&x&x&x&x & {\bf 9}\\
\hline
\multirow{3}{*}{2} & Precedence(Position patient, Anatomic identification) &\multirow{3}{*}{x}&\multirow{3}{*}{x}&\multirow{3}{*}{x}&\multirow{3}{*}{x}&\multirow{3}{*}{x}&\multirow{3}{*}{x}&\multirow{3}{*}{x}&\multirow{3}{*}{x}&\multirow{3}{*}{x}&\multirow{3}{*}{}
& \multirow{3}{*}{\bf 9}\\
 & Precedence(Position patient, Doppler identification) & & & & & & & & & & & \\
 & Precedence(Position patient, Compression identification) & & & & & & & &  & & &\\
\hline
\multirow{2}{*}{3} & Exclusive Choice 1~of~3 (Anatomic identification, & \multirow{2}{*}{x}&\multirow{2}{*}{x}&\multirow{2}{*}{x}&\multirow{2}{*}{}&\multirow{2}{*}{x}&\multirow{2}{*}{}&\multirow{2}{*}{x}&\multirow{2}{*}{x}&\multirow{2}{*}{x}&\multirow{2}{*}{x}
& \multirow{2}{*}{\bf 8}\\
& Doppler identification, Compression identification) & & & & & & & & & & & \\
\hline
4 & Exactly 1 (Anesthetize) & &&&&&&&&&& {\bf 0}\\
\hline
5 & Precedence(Anesthetize,Puncture) &&&&&&&&&& & {\bf 0}\\
\hline
6 & Alternate Response(Puncture,  Blood return) &&&&&&&&&& & {\bf 0}\\
\hline
7 & Precedence(Remove Trocar, Wire in good position) & &&&&&&&&&& {\bf 0}\\
\hline
8 & Precedence(Wire in good position, Advance catheter) & &&&&x&&&&x&& {\bf 2}\\
\hline
9 & Response(Remove guidewire, Check flow and reflow) &&&&&&&&&& & {\bf 0}\\
\hline
10 & Response(Install guidewire,  Remove guidewire) &&&&&&&&&& & {\bf 0}\\
\hline
\hline
& {\bf Total} & \bf{2}&\bf{3}&\bf{3}&\bf{2}&\bf{4}&\bf{2}&\bf{3}&\bf{3}&\bf{4}&\bf{2} & {\bf 28}\\
\hline
\hline
\end{tabular}
\end{table}
\subsubsection{Time Analysis}
\begin{figure}[!htb]
\centering
\includegraphics[width=0.99\textwidth]{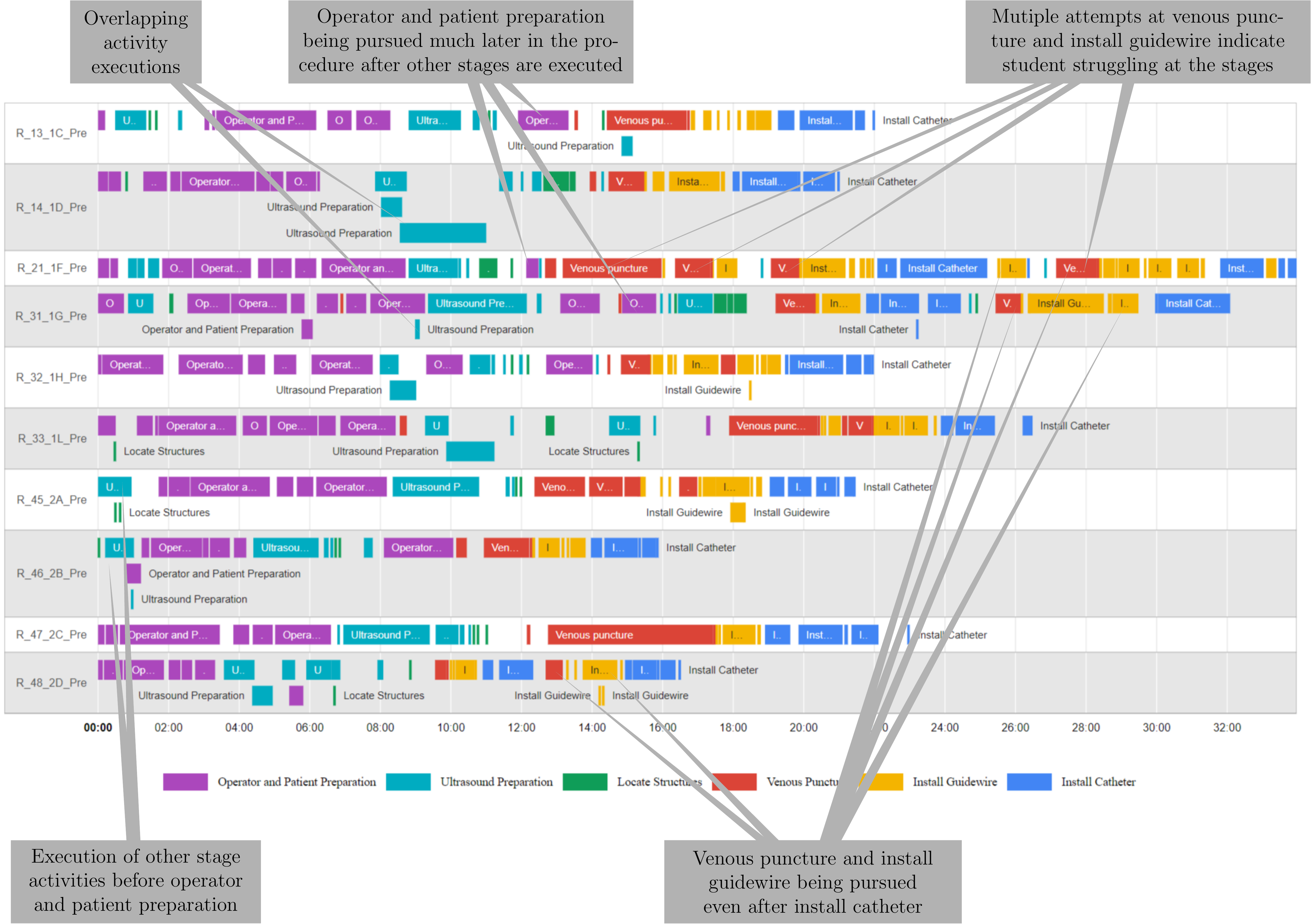}
\caption{Temporal view of the activities performed at the stage level by different students in the PRE round.}
\label{fig:stageAsActivityAllResourcesPreGTL}
\end{figure}
We analyzed the processing times of various activities and stages to see any patterns of behavior. While individual analysis will provide insights about how each student performed, cross-comparison across students provide different sets of insights. \figurename~\ref{fig:stageAsActivityAllResourcesPreGTL} depicts the temporal view of how each student performed during the procedure in the PRE round. We can see that student \texttt{\detokenize{R_21_1F}} took the longest time in performing the procedure ($\approx 34$ mins) while \texttt{\detokenize{R_46_2B}} is the fastest and completed the procedure in $\approx 16$ mins. It is important to note that with the exception of two students (\texttt{\detokenize{R_21_1F}} and \texttt{\detokenize{R_47_2C}}), all others exhibited overlapping execution of activities. Such overlapping execution is observed both of activities within the same stage (e.g., \texttt{\detokenize{R_14_1D}} has overlapping execution among activities of ultrasound preparation stage) and activities of different stages (e.g., \texttt{\detokenize{R_13_1C}} has ultrasound preparation activities overlapping with venous puncture). We could also see some patterns of abnormality, e.g., as per the protocol install catheter should be the last stage. However, three students \texttt{\detokenize{R_21_1F}}, \texttt{\detokenize{R_31_1G}}, and \texttt{\detokenize{R_48_2D}}, go back to previous stages (venous puncture, install guidewire, and in some cases, locate structures) even after install catheter stage has started.
\begin{figure}[!htb]
\centering
\includegraphics[width=0.99\textwidth]{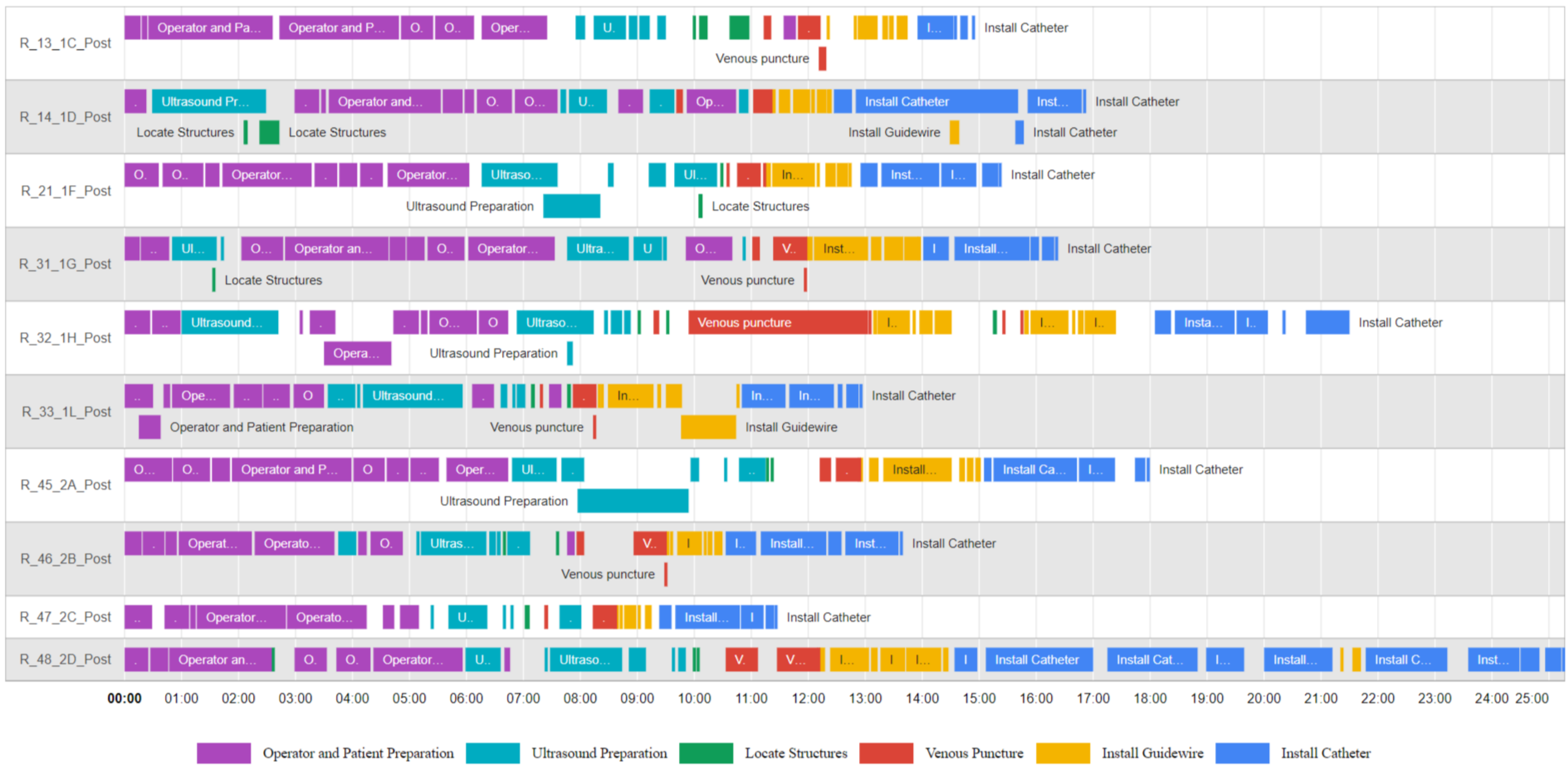}
\caption{Temporal view of the activities performed at the stage level by different students in the POST round.}
\label{fig:stageAsActivityAllResourcesPostGTL}
\end{figure}

\figurename~\ref{fig:stageAsActivityAllResourcesPostGTL} depicts the temporal view of how each student performed during the procedure in the POST round. We could see that the longest duration of executing the procedure in the POST round has reduced to $\approx 25$ mins. Overlapping execution of activities is still noticed in the POST round by $8$ students (\texttt{\detokenize{R_47_2C}} and \texttt{\detokenize{R_48_2D}} being the exceptions). However, in $6$ of these $8$ scenarios, the overlap is between activities belonging to the same stage. Only students \texttt{\detokenize{R_14_1D}} and \texttt{\detokenize{R_21_1F}} execute activities belonging to different stages simultaneously. Unlike the PRE round, once install catheter stage has started, previous stages were not revisited except for two students \texttt{\detokenize{R_14_1D}} and \texttt{\detokenize{R_48_2D}} and here too, only the immediate previous stage, i.e., install guidewire was revisited.
\subsubsection{Statistical Analysis}
A pertinent question to ask is {\it whether any improvement in student's performance has been noticed between the PRE and POST rounds?}. To answer this, we adopted the $t-$test statistics, which compares the means of two groups (the {\sf null} hypothesis states that the means of the two groups is the same, i.e., $H_0 : \mu_1 - \mu_2 = 0$). We apply the t-tests on both the control-flow conformance and time performance for the different approaches adopted in this paper. \tablename~\ref{tab:statisticalanalysisoverall} depicts the results of $t$-test on the control-flow deviations using the trace alignment and DECLARE analysis and the performance analysis on time (processing time and turnaround time) on the entire procedure. The values in the table specify the mean and standard deviation in the format $\mu (\sigma)$ and the p-value of the $t$-test. The $p-$value of all the aspects is $\le 0.05$ implying that the difference between the means is statistically significant and that the {\sf null} hypothesis has to be {\it rejected}. This clearly states that there is an improvement in student's performance in the POST round when compared to the PRE round in both the control-flow and time perspectives.
\begin{table}[!htb]
\caption{$t$-test analysis of control-flow conformance and the time perspective performance between the PRE and POST scenarios.}
\label{tab:statisticalanalysisoverall}
\centering
\begin{tabular}{p{0.2\textwidth}p{0.3\textwidth}p{0.2\textwidth}p{0.2\textwidth}r}
\hline
& & {\bf PRE} & {\bf POST} & {\bf p Value}\\
\hline
\multirow{2}{*}{\parbox{0.18\textwidth}{Control-flow Deviations} }& Trace Alignment  & $~~23.8~(10.45)$ & $~~12.9~(6.49)$ & ${\bf 0.011}$\\
\cline{2-5}
& DECLARE (LTL) & $~~~4.3~(1.34)$ & $~~~2.8~(0.79)$& ${\bf 0.007}$\\
\hline
\hline
\multirow{3}{*}{Time Analysis} & Turnaround time (tat) & $1405.0~(355.0)$ & $998.0~(248.0)$ & ${\bf 0.008}$\\
\cline{2-5}
& Processing time & $1062.0~(288.0)$ & $811.0~(199.0)$ & ${\bf 0.036}$ \\
\cline{2-5}
& \parbox{0.28\textwidth}{Processing time (as \% of tat)} & $~~~75.5~(~~5.2)$& $~~81.4~(~~5.9)$& ${\bf 0.030}$\\
\hline
\end{tabular}
\end{table}

\tablename~\ref{tab:statisticalanalysisstagelevel} depicts the $t-$test results on different stages on both the control-flow deviations (using trace alignment) and time perspectives (processing times). From the $p-$values of control-flow deviations, we can see that the {\sf null} hypothesis is outright rejected for the stages ultrasound preparation, venous puncture, and install guidewire ($p-$value $\le 0.05$), implying that there is a significant difference in the deviations pertaining to these three stages between the PRE and POST round (with the POST round being more conformant). However, such an improvement is not eminent in the other three stages, viz., operator and patient preparation, locate structures, and install catheter. {\bf This indicates that the students need better education/practice in these three stages}. From the time perspective, one could notice significant improvement in the POST round for the venous puncture stage, while somewhat improvement is found in the ultrasound preparation and install guidewire stages too.
\begin{table}[!htb]
\caption{$t$-test analysis of control-flow conformance and the time perspective performance between the PRE and POST scenarios at stage level.}
\label{tab:statisticalanalysisstagelevel}
\centering
\begin{tabular}{p{0.3\textwidth}p{0.3\textwidth}p{0.15\textwidth}p{0.15\textwidth}r}
\hline
& & {\bf PRE} & {\bf POST} & {\bf p Value}\\
\hline
\multirow{6}{*}{\parbox{0.28\textwidth}{Stage Trace Alignment Deviations}} & {\parbox{0.28\textwidth}{Operator and Patient Preparation}} & $4.40~(0.84)$ & $4.10~(1.37)$ & $0.56$\\
\cline{2-5}
& {\parbox{0.38\textwidth}{Ultrasound Preparation}} & $2.70~(1.16)$ & $1.40~(0.84)$ & ${\bf 0.01}$ \\
\cline{2-5}
& {\parbox{0.38\textwidth}{Locate Structures}} & $2.40~(1.26)$ & $2.70~(1.34)$ & $0.61$ \\
\cline{2-5}
& {\parbox{0.38\textwidth}{Venous Puncture}} & $2.20~(2.39)$ & $0.20~(0.63)$ & ${\bf 0.02}$ \\
\cline{2-5}
& {\parbox{0.38\textwidth}{Install Guidewire}} & $7.50~(4.77)$ & $2.10~(2.28)$ & ${\bf 0.005}$ \\
\cline{2-5}
& {\parbox{0.38\textwidth}{Install Catheter}} & $1.80~(2.62)$ & $0.80~(1.62)$ & $0.32$ \\
\cline{2-5}
\hline
\hline
\multirow{6}{*}{\parbox{0.28\textwidth}{Stage Processing Time}} & {\parbox{0.28\textwidth}{Operator and Patient Preparation}} & $388.0~(106.0)$ & $331~(~67.0)$ & $0.16$\\
\cline{2-5}
& {\parbox{0.38\textwidth}{Ultrasound Preparation}} & $197.0~(~52.0)$ & $155~(~57.0)$ & $0.10$ \\
\cline{2-5}
& {\parbox{0.38\textwidth}{Locate Structures}} & $~17.9~(~24.2)$ & $~~6~(~10.8)$ & $0.17$ \\
\cline{2-5}
& {\parbox{0.38\textwidth}{Venous Puncture}} & $160.2~(102.1)$ & $~54.6~(~52.2)$ & ${\bf 0.009}$ \\
\cline{2-5}
& {\parbox{0.38\textwidth}{Install Guidewire}} & $135.3~(~89.1)$ & $~83.8~(~43.4)$ & $0.117$ \\
\cline{2-5}
& {\parbox{0.38\textwidth}{Install Catheter}} & $163.3~(~62.3)$ & $180.1~(131.7)$ & $0.71$ \\
\cline{2-5}
\hline
\end{tabular}
\end{table}
\section{Conclusions} \label{sec:conclusions}
In this paper, we proposed an approach to analyze the event log pertaining to medical training process of students studying CVC procedure using ultrasound. We have studied the control-flow conformance and temporal performance. The key findings are as follows:
\begin{itemize}
\item although students exhibit more compliance when looking at activity executions of each stage separately, they tend to struggle with the overall procedure. There is a greater tendency of interleaved execution of activities from different stages
\item we noticed a tendency of performing some activities concurrently (overlapping executions), if undesirable from a process point of view, the students need to be trained on this
\item students perform significantly better in terms of compliance to the protocol and processing times during their POST rounds when compared to their PRE rounds at the overall procedure level. However, not much improvement w.r.t the control-flow deviations has been observed in the operator and patient preparation, locate structures, and install catheter stages.
\item a consistent issue that is observed is w.r.t the two exclusive choice constructs, one in the locate structures stage and the other in the install guidewire stage. Students almost always execute at least two activities while the protocol says one of them to suffice.
\item even after POST round, the following activities need special attention
\begin{itemize}
\item Prepare implements, hand washing and Get sterile clothes are the most deviant activities in the operator and patient preparation stage
\item Ultrasound configuration is the most deviant activity in the ultrasound preparation stage
\item if there is one stage that almost all students flunk, it is the locate structures stage
\item Widen pathway and advance catheter seem to require attention
\end{itemize}
\end{itemize}
The techniques adopted in our approach provide easily interpretable and actionable insights even to non process mining experts. We conjecture a checklist adoption in teaching or practising phase might benefit the students in learning the skills to perfection.
\bibliographystyle{splncs04}
\bibliography{CCC2019}

\appendix
\section{Trace Alignment of Individual Students}\label{sec:appendtracealign}
\subsection{\texttt{\detokenize{R_14_1D}}}
\begin{figure}[H]
\centering
\includegraphics[width=0.99\textwidth]{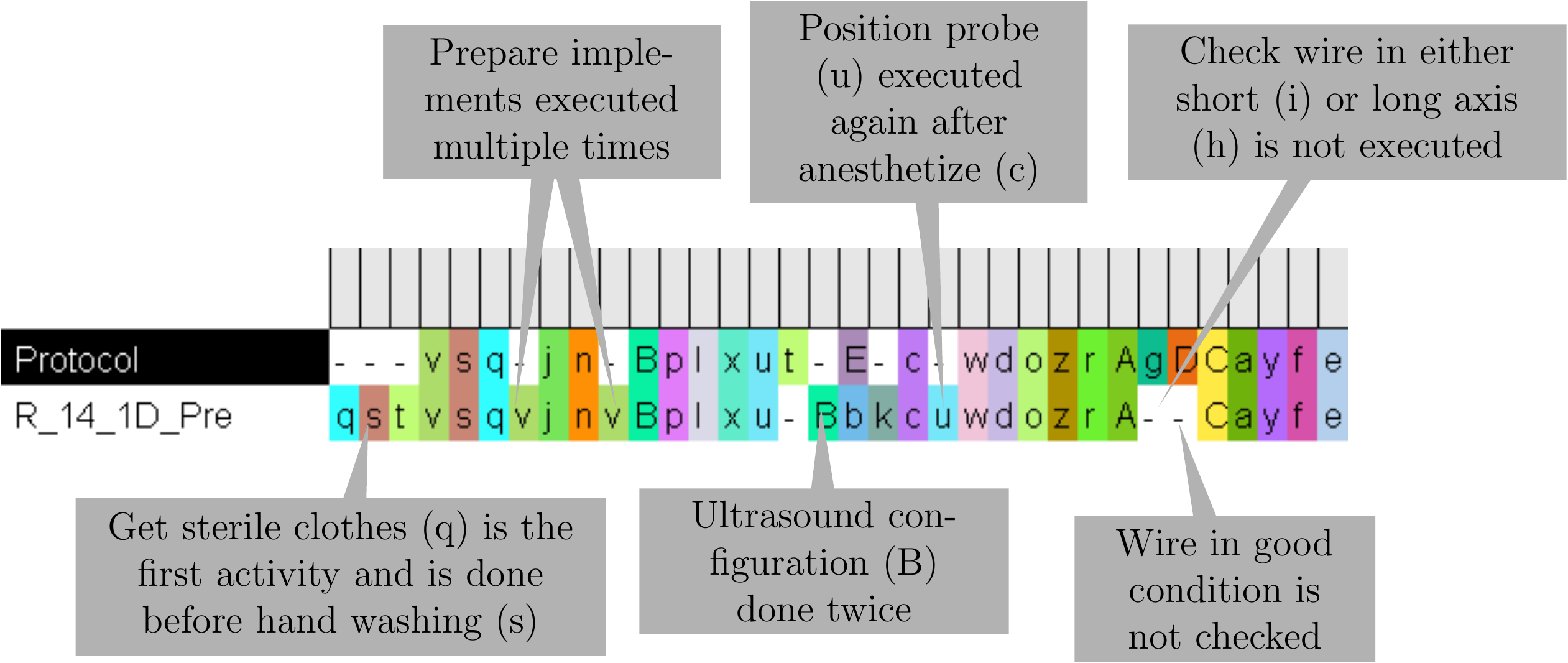}
\caption{Trace alignment of PRE round and protocol for student \texttt{\detokenize{R_14_1D}}.}
\end{figure}
\begin{figure}[H]
\centering
\includegraphics[width=0.99\textwidth]{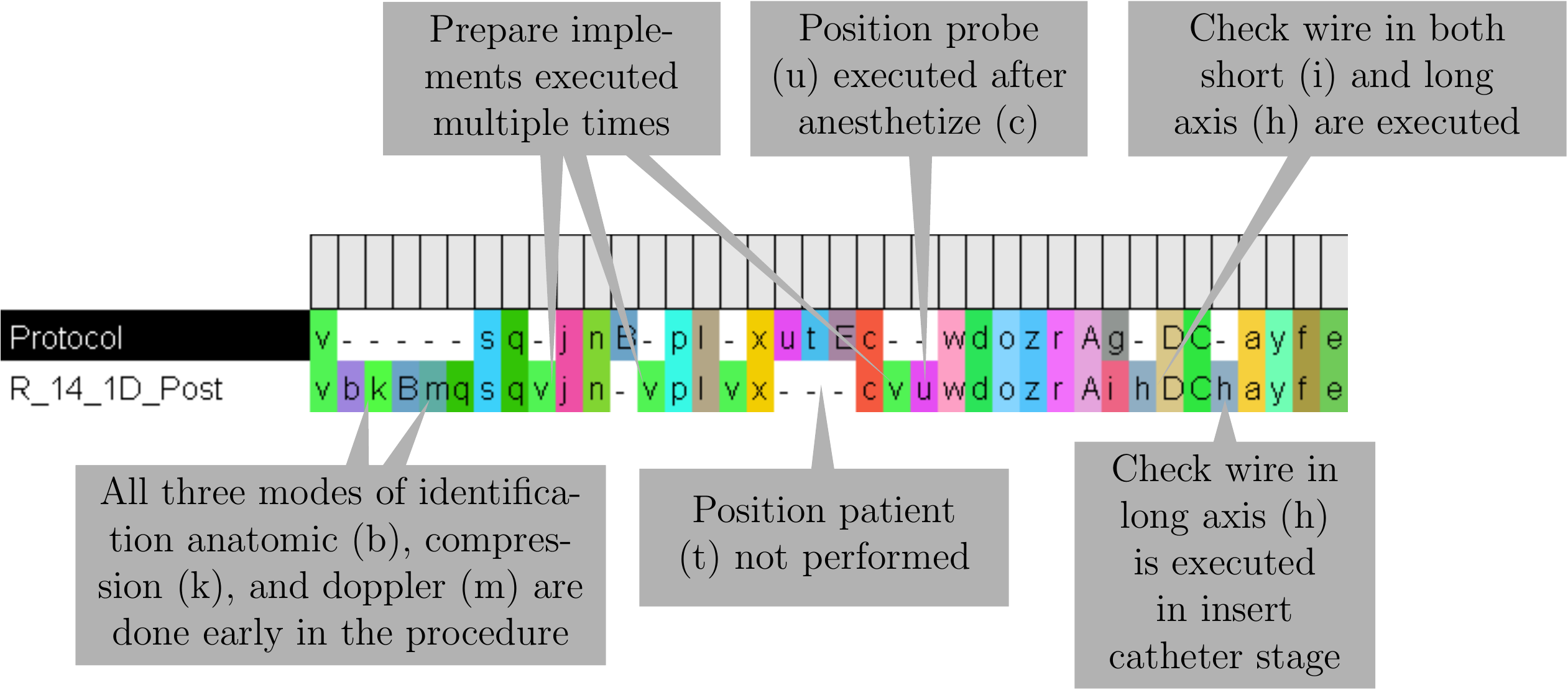}
\caption{Trace alignment of POST round and protocol for student \texttt{\detokenize{R_14_1D}}.}
\end{figure}
\subsection{\texttt{\detokenize{R_21_1F}}}
\begin{figure}[H]
\centering
\includegraphics[width=0.99\textwidth]{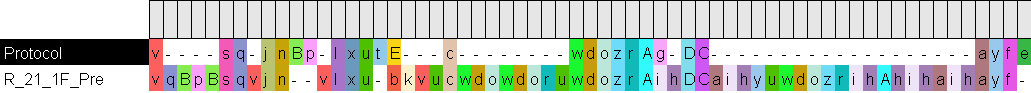}
\caption{Trace alignment of PRE round and protocol for student \texttt{\detokenize{R_21_1F}}.}
\label{fig:R_21_1FPreTraceAlignment}
\end{figure}
\begin{figure}[H]
\centering
\includegraphics[width=0.99\textwidth]{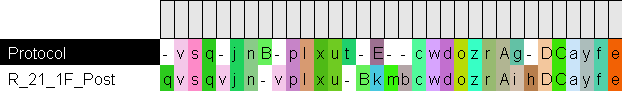}
\caption{Trace alignment of POST round and protocol for student \texttt{\detokenize{R_21_1F}}.}
\end{figure}
\subsection{\texttt{\detokenize{R_31_1G}}}
\begin{figure}[H]
\centering
\includegraphics[width=0.99\textwidth]{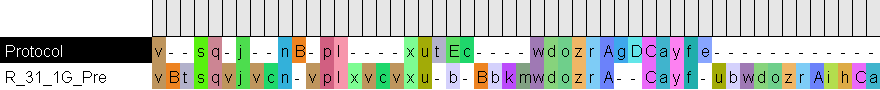}
\caption{Trace alignment of PRE round and protocol for student \texttt{\detokenize{R_31_1G}}.}
\end{figure}
\begin{figure}[H]
\centering
\includegraphics[width=0.99\textwidth]{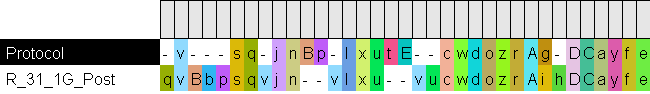}
\caption{Trace alignment of POST round and protocol for student \texttt{\detokenize{R_31_1G}}.}
\end{figure}
\subsection{\texttt{\detokenize{R_32_1H}}}
\begin{figure}[H]
\centering
\includegraphics[width=0.99\textwidth]{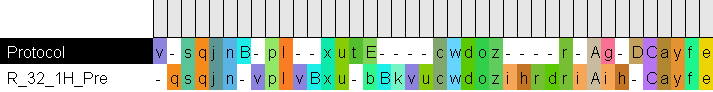}
\caption{Trace alignment of PRE round and protocol for student \texttt{\detokenize{R_32_1H}}.}
\end{figure}
\begin{figure}[H]
\centering
\includegraphics[width=0.99\textwidth]{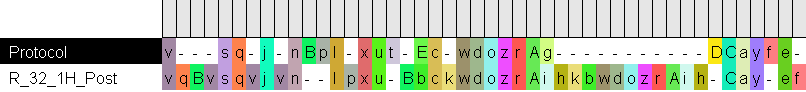}
\caption{Trace alignment of POST round and protocol for student \texttt{\detokenize{R_32_1H}}.}
\end{figure}
\subsection{\texttt{\detokenize{R_33_1L}}}
\begin{figure}[H]
\centering
\includegraphics[width=0.99\textwidth]{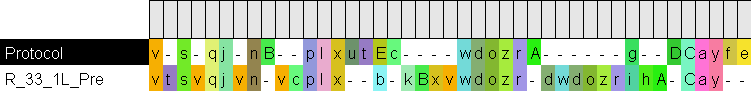}
\caption{Trace alignment of PRE round and protocol for student \texttt{\detokenize{R_33_1L}}.}
\end{figure}
\begin{figure}[H]
\centering
\includegraphics[width=0.99\textwidth]{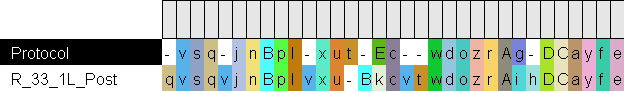}
\caption{Trace alignment of POST round and protocol for student \texttt{\detokenize{R_33_1L}}.}
\end{figure}
\subsection{\texttt{\detokenize{R_45_2A}}}
\begin{figure}[H]
\centering
\includegraphics[width=0.99\textwidth]{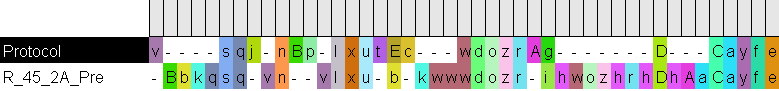}
\caption{Trace alignment of PRE round and protocol for student \texttt{\detokenize{R_45_2A}}.}
\end{figure}
\begin{figure}[H]
\centering
\includegraphics[width=0.99\textwidth]{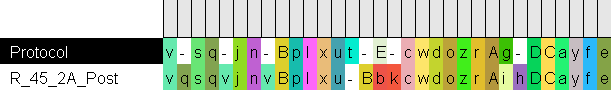}
\caption{Trace alignment of POST round and protocol for student \texttt{\detokenize{R_45_2A}}.}
\end{figure}
\subsection{\texttt{\detokenize{R_46_2B}}}
\begin{figure}[H]
\centering
\includegraphics[width=0.99\textwidth]{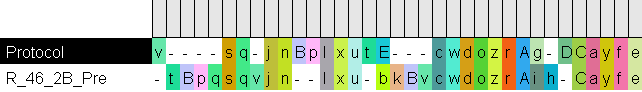}
\caption{Trace alignment of PRE round and protocol for student \texttt{\detokenize{R_46_2B}}.}
\end{figure}
\begin{figure}[H]
\centering
\includegraphics[width=0.99\textwidth]{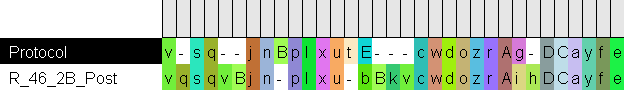}
\caption{Trace alignment of POST round and protocol for student \texttt{\detokenize{R_46_2B}}.}
\end{figure}
\subsection{\texttt{\detokenize{R_47_2C}}}
\begin{figure}[H]
\centering
\includegraphics[width=0.99\textwidth]{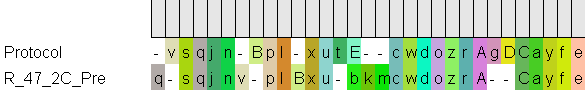}
\caption{Trace alignment of PRE round and protocol for student \texttt{\detokenize{R_47_2C}}.}
\end{figure}
\begin{figure}[H]
\centering
\includegraphics[width=0.99\textwidth]{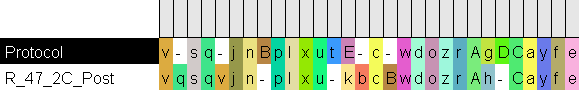}
\caption{Trace alignment of POST round and protocol for student \texttt{\detokenize{R_47_2C}}.}
\end{figure}
\subsection{\texttt{\detokenize{R_48_2D}}}
\begin{figure}[H]
\centering
\includegraphics[width=0.99\textwidth]{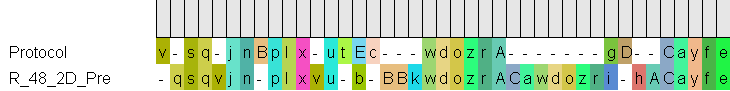}
\caption{Trace alignment of PRE round and protocol for student \texttt{\detokenize{R_48_2D}}.}
\end{figure}
\begin{figure}[H]
\centering
\includegraphics[width=0.99\textwidth]{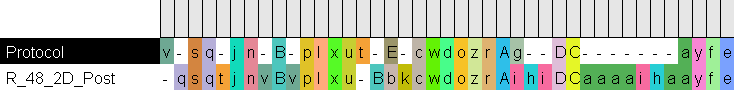}
\caption{Trace alignment of POST round and protocol for student \texttt{\detokenize{R_48_2D}}.}
\end{figure}

\end{document}